\documentclass[traditabstract]{aa} 
\providecommand{\reff}{}  

\usepackage{graphicx}
\usepackage{natbib}
\usepackage{mathabx}
\usepackage{deluxetable}
\usepackage[pdftex,                
bookmarks=true,                    
bookmarksnumbered=true,            
colorlinks=true,            
citecolor=blue,         
linkcolor=blue,     
menucolor=blue,         
urlcolor=cyan,       
linkbordercolor={0 0 1},           
pdfborder= {0 0 1},
frenchlinks=True]{hyperref}
\usepackage{txfonts}
\providecommand{\eprint}[1]{\href{http://arxiv.org/abs/#1}{#1}}
\providecommand{\adsurl}[1]{\href{#1}{ADS}}

\newcommand{\fig}[5]{
        \begin{figure}[!bt]
        \begin{centering}
        \includegraphics[#3]{#2}
      \end{centering}
      \renewcommand{\baselinestretch}{1}
        \vspace*{-.3in}
        \caption[#4]{#5}
        \label{fig:#1}
        \end{figure}}

\def\micron{\ensuremath{\mu m}}

\newcommand{\gjb}{GJ~3470b}
\newcommand{\gj}{GJ~3470}
\def\rprs{\ensuremath{0.0789^{+0.0021}_{-0.0019}}}

\def\porb{\ensuremath{3.336665 \pm 0.000008}~d}

\begin{document}

\title{Warm Ice Giant GJ~3470b. I. A Flat Transmission Spectrum Indicates a Hazy, Low-Methane, and/or Metal-rich Atmosphere}

\authorrunning{Crossfield et al.}
\titlerunning{GJ~3470b has a Hazy, Low-Methane, or Metal-rich Atmosphere}

   \author{Ian J. M. Crossfield\inst{1}
     Travis Barman\inst{2}
     Brad M. S. Hansen\inst{3}
     Andrew W. Howard\inst{4}
             }

          \institute{
             $^1$~Max-Planck Institut f\"ur Astronomie, K\"onigstuhl 17, 69117, Heidelberg, Germany.  \email{\href{mailto:ianc@mpia.de}{ianc@mpia.de}} \\
$^2$~Lowell Observatory, 1400 West Mars Hill Road, Flagstaff, AZ 86001, USA \\
$^3$~Department of Physics \& Astronomy, University of California Los Angeles, Los Angeles, CA 90095, USA\\
$^4$~Institute for Astronomy, University of Hawaii, 2680 Woodlawn Drive, Honolulu, HI 96822, USA
            }

   \date{A\&A accepted, 2013/08/29. Light curves available online.}

 
  \abstract
  { We report our spectroscopic investigation of the transiting ice
    giant \gjb's atmospheric transmission, and the first results of
    extrasolar planet observations from the new Keck/MOSFIRE
    spectrograph.  We measure a planet/star radius ratio of \rprs\ in
    a bandpass from 2.09--2.36\,\micron\ and in six narrower bands
    across this wavelength range. When combined with existing
    broadband photometry, these measurements rule out cloud-free
    atmospheres in chemical equilibrium assuming either solar
    abundances (5.4$\sigma$ confidence) {\reff or} a moderate level of
    metal enrichment (50$\times$ solar abundances, 3.8$\sigma$),
    confirming previous results that such models are not
    representative for cool, low-mass, externally irradiated
    extrasolar planets.  Current measurements are consistent with a
    flat transmission spectrum, which suggests that the atmosphere is
    explained by high-altitude clouds and haze, disequilibrium
    chemistry, unexpected abundance patterns, or the atmosphere is
    extremely metal-rich ($\gtrsim200\times$ solar). {\reff Because
      GJ~3470b's low bulk density sets an upper limit on the planet's
      atmospheric enrichment of $\lesssim300\times$ solar, the
      atmospheric mean molecular weight must be $\lesssim 9$. Thus, if
      the atmosphere is} cloud-free its spectral features should be
    detectable with future observations. Transit observations at
    shorter wavelengths will provide the best opportunity to
    discriminate between plausible scenarios. We obtained optical
    spectroscopy with the GMOS spectrograph, but these observations
    exhibit large systematic uncertainties owing to thin, persistent
    cirrus conditions. Finally, we also provide the first detailed
    look at the steps necessary for well-calibrated MOSFIRE
    observations, and provide advice for future observations with this
    instrument.  }
   { }
   { }
   { }
   {}


   \maketitle
%

\section{Introduction}

\subsection{Sub-Jovian Planets are Common}

Planets with masses similar to Uranus and Neptune are a plausible link
between gaseous Jupiter-class giants and rocky terrestrial-class
planets, and are often defined by their failure to capture gas from
the protoplanetary nebula and become fully fledged gas giants. Yet
their presence and properties, by virtue of that very same lack of a
natural explanation, offer crucial information about how planets form
and possibly move during the evolution of planetary systems.

Scenarios to explain the origins of hot Jupiters by migration predicted
that Neptune-mass planets would be scarce at short periods, because
of the fine tuning required to avoid the rapid
accretion of gas from the nebula while also using the gas to migrate
inwards \citep{ida:2008,mordasini:2009b}. Yet radial
velocity searches have uncovered many planets which fall into
this class \citep{howard:2010,mayor:2011}. Thus, 
in-depth study of  these enigmatic planets could potentially
shed new light on the nature of what now appears to one of
the more common class of planetary system.

Although hot Neptunes are not expected from the most commonly adopted
migration models, several other possibilities have been
discussed. Tidal capture has been proposed as an explanation for hot
Jupiters \citep{rasio:1996,fabrycky:2007,nagasawa:2008,naoz:2011} and
could potentially produce low mass planets, but the effectiveness of
this process depends strongly on the poorly constrained efficiency of
tidal dissipation in such planets
\citep[e.g.,][]{kennedy:2008,barnes:2009}.  Migration of several low-mass
planets in a resonant chain may help to slow the migration and prevent
the planets from being swallowed by the star
\citep{terquem:2007,ogihara:2009,ida:2010}. Neptune-mass planets may
also represent the accretion-starved analogues of Jupiter-class
planets that assemble in situ \citep{hansen:2012}, although this
likely requires a substantial inventory of solid material to assemble
a sufficiently large core mass.

All of these scenarios imply different histories and compositions for
the resulting planets, and this can potentially be probed by an in
depth study of the planetary atmosphere and composition.  Measurements
of bulk properties such as planetary radii and masses provide insight
into planetary interior structures and compositions. This approach is
powerful, but for planets with Uranus-like bulk properties it long
been known that degeneracies exist and the interior structure and
composition cannot be uniquely inferred \citep[e.g.,][and references
therein]{wildt:1947, miner:1990}. The same limitation apply to
extrasolar planets
\citep[e.g.,][]{adams:2008,figueira:2009,nettelmann:2010}. A
complementary technique is to observe a planet's atmosphere (whether
by transits, occultations, phase curves, in-situ probes, etc.),
thereby constraining the atmosphere's structure and chemical
composition and (we hope) inferring something about the planet's
interior composition. Among other goals, such measurements should
provide clues about the planet's formation history which are encoded
in their bulk and atmospheric properties
\citep[e.g.,][]{oberg:2011,fortney:2013}.

\subsection{Atmospheric Characterization of Short-Period Exoplanets}
Since the first detection of an exoplanet's atmosphere
\citep{charbonneau:2002} numerous hot Jupiters have been studied via
the emission or transmission of their atmospheres. This has led to
robust detections of numerous atomic species and molecules
\citep{charbonneau:2002,barman:2007}, high-altitude hazes
\citep{pont:2008,pont:2013,lecavelier:2008haze189}, and a diverse
range of atmospheric circulation patterns
\citep{knutson:2007b,crossfield:2010} and albedos
\citep{rowe:2008,demory:2011}. These discoveries continue to fuel an
ongoing revolution in the field of externally irradiated planetary
atmospheres.  However, despite the greater frequency with which
smaller, lower-mass planets occur most are not amenable to such
followup observations: their transit depths, temperatures, and/or host
star apparent fluxes are typically too low.

To date only two planets of roughly Neptune size or smaller have been
subjected to detailed scrutiny: GJ~1214b and GJ~436b. The small, cool
planet GJ~1214b has {\reff one of} the most complete transmission spectra of any
extrasolar planet \citep[][and references
therein]{bean:2011,berta:2012}, but these measurements are consistent
with a flat, featureless spectrum. GJ~1214b's bulk properties likely
require a substantial volatile envelope, but this envelope's
composition is unknown \citep{nettelmann:2010,rogers:2010}. The
ensemble of current data indicates that the planet's atmosphere is
either shrouded in an opaque haze  or it is
composed predominately of heavy molecules such as H$_2$O
\citep[][]{howe:2012,morley:2013,benneke:2013}.

Photometry of the hotter, more massive GJ 436b has been more
revealing: the planet's substantial H$_2$ envelope
\citep{adams:2008,figueira:2009} is significantly depleted in CH$_4$
and enhanced in CO when compared to equilibrium conditions and solar
composition \citep{stevenson:2010,knutson:2011}. These results suggest
that the planet's atmosphere is metal-rich compared to its host star,
and may exhibit strong internal diffusion that brings CO up into the
observable photosphere and/or significant photochemistry
\citep{line:2011,madhusudhan:2011}.  Recent investigations of very
high metallicity atmospheres ($>100\times$ solar) indicate that such
compositions can also explain observations of GJ~436b
\citep{fortney:2013,moses:2013}.  However, to date no reliable
spectroscopy has been obtained for GJ~436b \citep[but
see][]{pont:2009,gibson:2011}.

\subsection{Introducing GJ~3470b}
The recently discovered planet GJ~3470b presents another excellent
target for atmospheric characterization of a relatively small and cool
object via transmission spectroscopy.  The planet was discovered by
radial velocity measurements and subsequently seen to transit
\citep{bonfils:2012}; subsequent observations demonstrate that the
planet is larger and of slightly lower mass than Uranus, with radius
4.8\,$R_\Earth$ and mass 14\,$M_\Earth$
\citep{demory:2013,fukui:2013,biddle:2013}. The planet orbits an early
M star whose mass and radius are roughly half that of the Sun, and
which may be slightly metal-rich
\citep{demory:2013,pineda:2013,biddle:2013}. The stellar flux varies
by $\sim$1\% in R band \citep{biddle:2013}, which will be important
for future work but is not significant for our relatively poor final
precision. We summarize the stellar and planetary parameters used in
this study in Table~\ref{tab:syspar}.

Compared to previously studied planets, \gjb's mass and equilibrium
temperature (600--800~K) lie between those of GJ~1214b and GJ~436b,
and \gjb\ is the largest of these three planets. According to interior
models, GJ~3470b must have a H$_2$ envelope that is $\sim$10\% the
mass of the planet \citep{fortney:2007,rogers:2010,
  valencia:2011,demory:2013}.  Thus \gjb, like GJ~436b, likely has a
H$_2$ envelope over a denser central core containing a large large ice
(H$_2$O, CH$_4$, NH$_3$) complement.

If \gjb\ has a clear atmosphere it presents an attractive target for
transmission spectroscopy owing to the planet's large atmospheric
scale height, transit depth and duration, and host star apparent
magnitude \citep{miller-ricci:2009}. In this case the high-altitude
opacity structure should be detectable as a variation of transit depth
with wavelength: i.e., the planet would appear larger at wavelengths
of high opacity.  A marginal detection of this
effect in broadband photometry has already been reported by
\cite{fukui:2013}.

However, optically thick clouds or haze can obscure atmospheric
features during transit and produce a flat transmission spectrum, and
haze has been invoked for several relatively cool (sub-1200~K)
transiting planets
\citep{knutson:2011,benneke:2012,pont:2013,gibson:2013,morley:2013}. The
most well-studied case is that of HD~189733b, in which the optically
thick haze first seen at visible wavelengths
\citep{pont:2008,lecavelier:2008haze189} appears to extend at least
into the near-infrared \citep{sing:2009haze}. Though the composition
of any such haze in these planets' atmospheres remains unknown, if
present a haze would substantially modify the emission and
transmission spectra by increasing the atmospheric opacity and thereby
raising the altitude of the effective photosphere and obscuring
signatures at deeper altitudes \citep{pont:2013}. Clouds have been
invoked to explain a range of observed characteristics of brown dwarf
atmospheres and self-luminous exoplanets
\citep{barman:2011,witte:2011,marley:2012}. As yet few such studies
have been undertaken for short-period exoplanets
\citep{zahnle:2009,morley:2013,parmentier:2013}; the conditions under
which such hazes may form in these objects' atmospheres (like their
interior compositions) remains unknown.

\subsection{Paper Overview}
We have undertaken a campaign of transit photometry and spectroscopy
to probe the physical parameters of the \gj\ system and the
atmospheric composition of \gjb. Paper~II \citep{biddle:2013} will 
discuss an analysis of new transit photometry which dramatically
improves the system parameters. In this work, we discuss our optical
and near-infrared (NIR) transmission spectroscopy obtained during two
transits of \gjb, which provides evidence for a flat transmission
spectrum. In \S~\ref{sec:data} we describe our observations and
initial data calibration.  In \S~\ref{sec:analysis} we describe our
approach to modeling light curves, accounting for limb darkening, and
estimating our final measurement uncertainties. In \S~\ref{sec:model}
we present our atmospheric models of \gjb's atmosphere, which leads to
our primary results: we rule out a solar-abundance atmosphere in
chemical equilibrium, and find that more metal-enriched atmospheres
are also disfavored. In \S~\ref{sec:disc} we discuss the implications
of these results in the broader context of atmospheric
characterization of cool, low-mass planets, and conclude in
\S~\ref{sec:conclusion}. Finally, in Appendix~\ref{sec:mosfire} we
present guidelines for improved calibration of observations with the
Keck Observatory's new MOSFIRE instrument.

\section{Data Acquisition and Calibration}
\label{sec:data}

\subsection{Keck/MOSFIRE}
\subsubsection{Observations}
We observed \gj\ on UT 2013-04-08 with the Keck~I telescope using the
new MOSFIRE multi-object spectrograph
\citep{mclean:2008,mclean:2010,mclean:2012,kulas:2012}. MOSFIRE is a
near-infrared multi-object spectrograph located on a rotatable mount
at the Keck~I Cassegrain focus. The instrument provides spectral
resolution of roughly 3500 (with a slit width of 0.7'') for targets
over a field of view of roughly 6'$\times$6'; a single photometric
band is covered at each instrument setting.  We have learned a
considerable amount about how best to obtain high-precision
spectrophotometric light curves with this instrument; we present the
finer points of this discussion in
Appendix~\ref{sec:mosfire}.

In total we obtained 401 K-band frames during 257~min and over an
airmass range of 1.00--2.3.  Because of \gj's bright infrared flux
\citep[$K_S=7.99$;][]{skrutskie:2006}, we used exposures of 5.82~s
duration with 2 coadds; each coadd used four non-destructive reads to
``sample up the ramp'' during the integration. We attempted to save
all non-destructive reads from the detector
(\citeauthor{bean:2011}~\citeyear{bean:2011} discuss why this is
desirable), but the instrument electronics were unable to keep
up with our high data rate and we were forced to abandon this
strategy.

\begin{figure}[tb!]
\centering
\includegraphics[width=9cm]{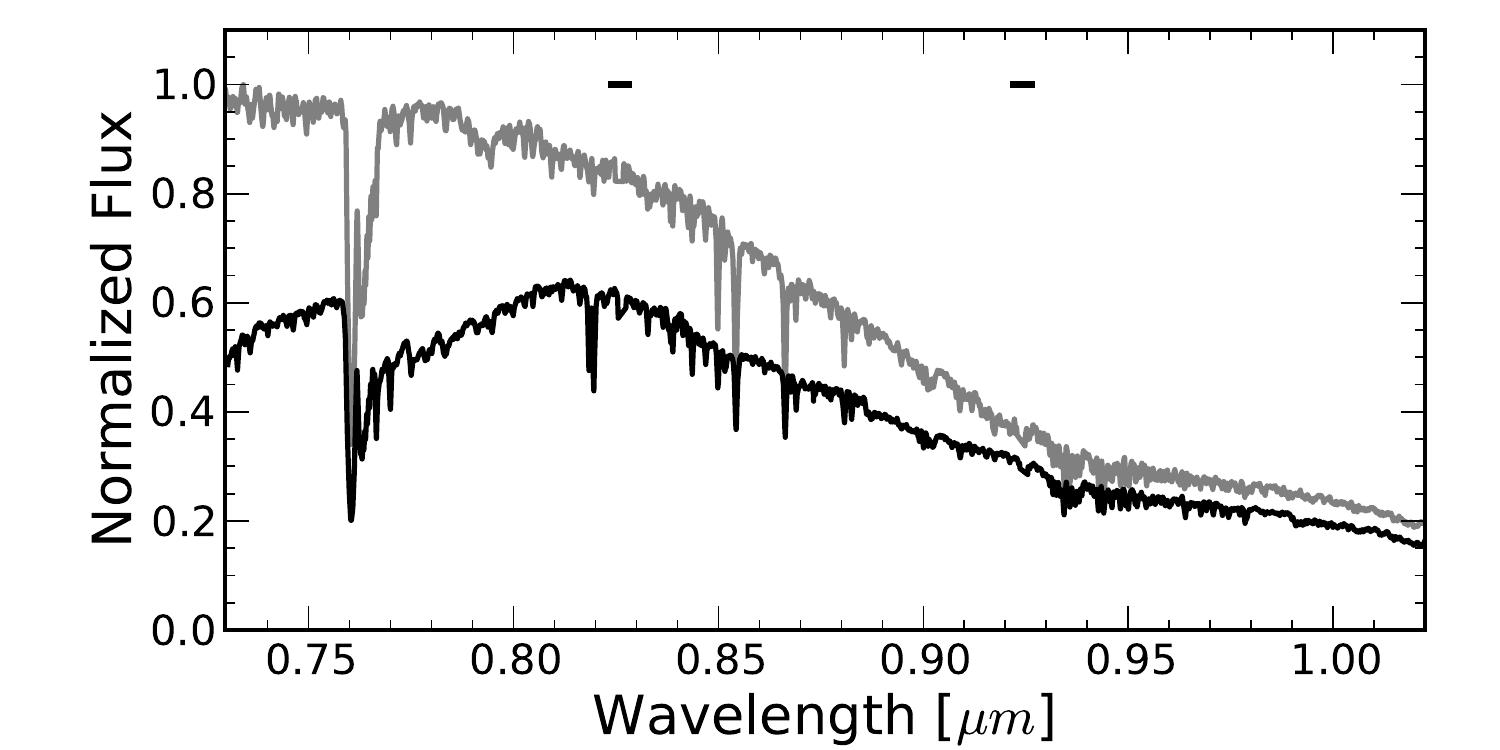}
\includegraphics[width=9cm]{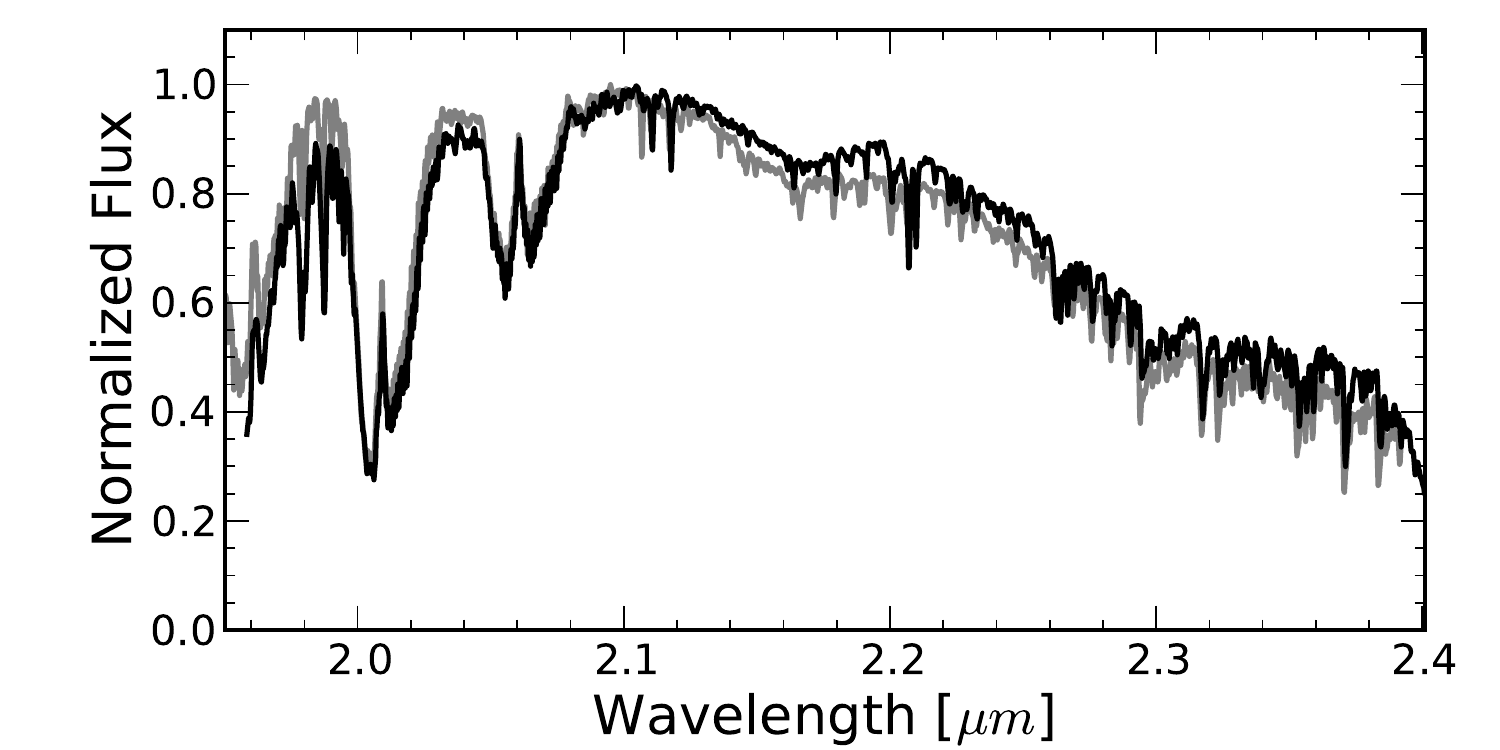}
\vspace{-0.2cm}
\caption{\footnotesize \label{fig:rawspec} GMOS (top) and MOSFIRE (bottom) spectra
  of exoplanet host star \gj\ (black) and simultaneously-observed
  comparison star TYC 1363-2087-1 (gray). The former is a
  main-sequence M dwarf, while the latter is a rather hotter giant.
  The short black bars in the top panel indicate the locations of the
  gaps between the GMOS CCDs. }
\end{figure}

We obtained simultaneous spectroscopy of GJ~3470, TYC 1363-2087-1, and
the fainter star 2MASS~07591321+1524069. Fig.~\ref{fig:rawspec} shows
the raw spectra of the first two stars. We discuss all objects'
spectral characteristics in \S~\ref{sec:comparison}, but we use only
the first two objects in our MOSFIRE analysis. We used wide (10'')
slits to ensure that the spectrograph captures essentially all stellar
flux, regardless of guiding errors or changes in seeing. We nodded the
telescope along the slit axis (see \S~\ref{sec:fringing}). Our spectra
cover wavelengths from 1.96--2.39\,\micron\ for both \gj\ and our
primary comparison star, and we achieved S/N of roughly
80--120~column$^{-1}$~frame$^{-1}$.

\begin{figure}[tb!]
\begin{minipage}{0.5\textwidth}
\centering
\includegraphics[width=4.5cm]{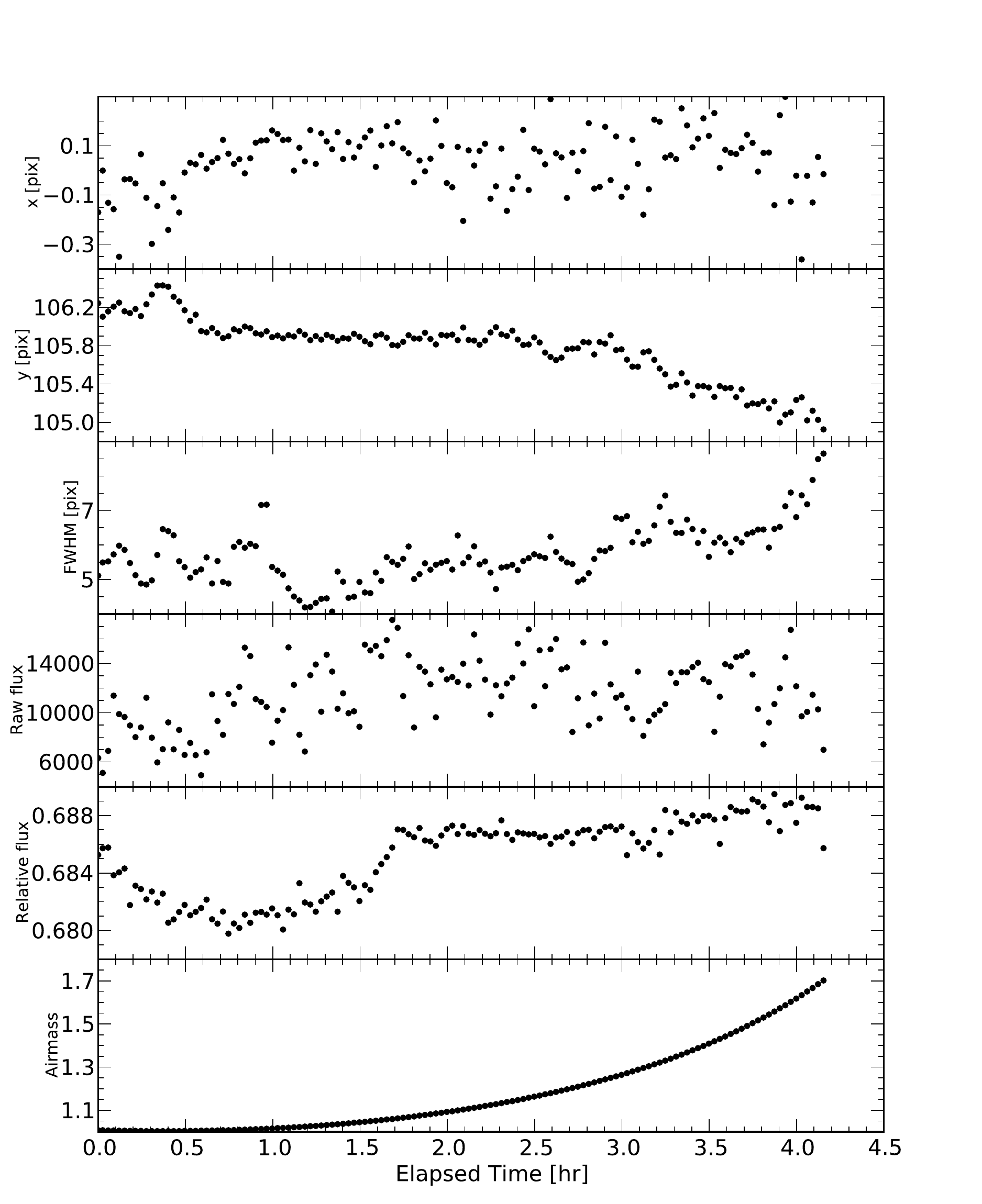}
\includegraphics[width=4.5cm]{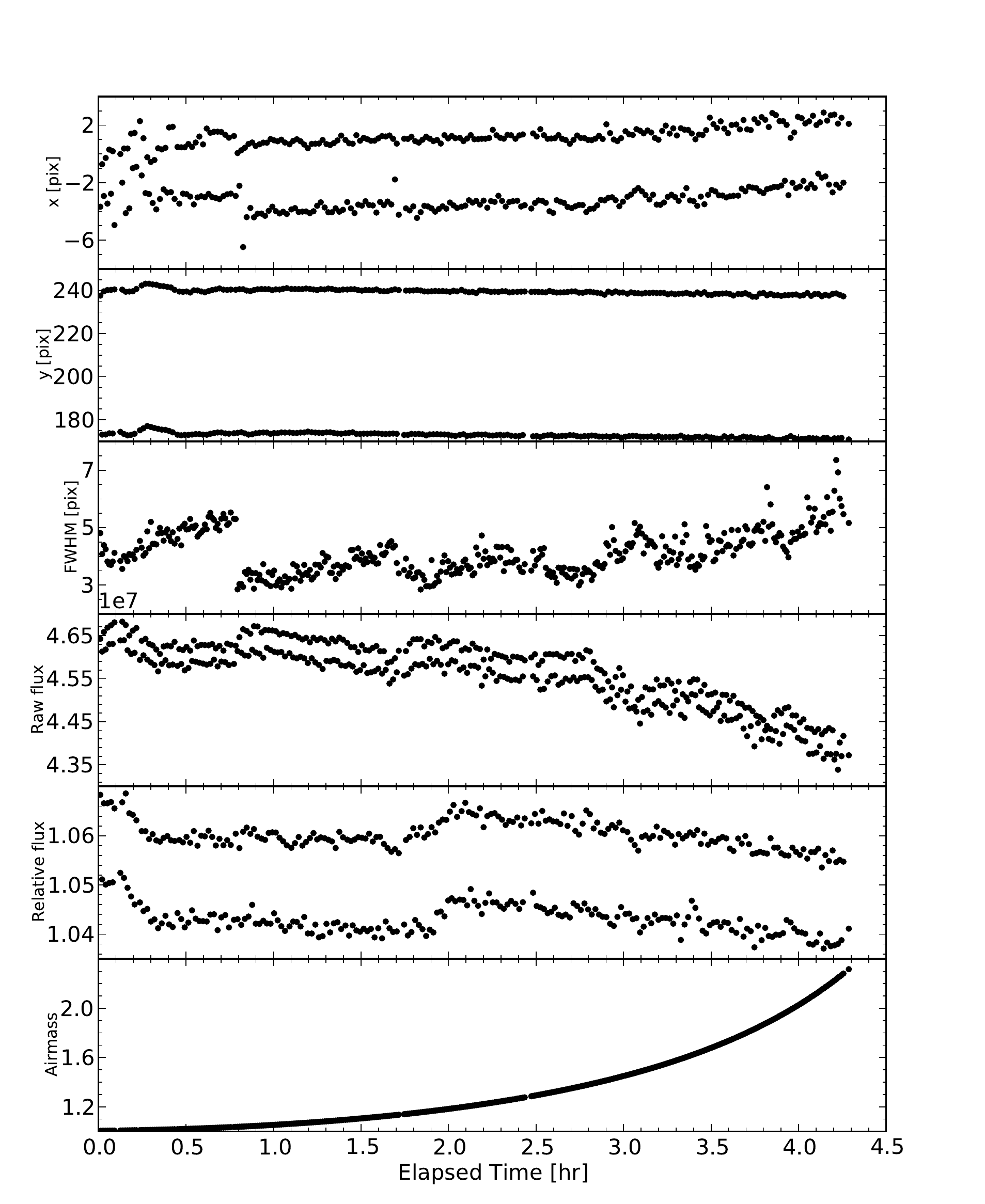}
\vspace{-0.2cm}
\end{minipage}
\caption{\footnotesize \label{fig:obs} Instrumental trends for spectroscopic
  observations of GJ~3470. {\em Left}: GMOS. The effects of
  intermittent cirrus are apparent in the raw stellar flux; this also
  increases the scatter in the relative light curve.  {\em Right}:
  MOSFIRE. Some parameters appear double-valued because data were
  taken while alternating between two nod positions on the slit. }
\end{figure}

We acquired \gj\ just as transit began. When observations were
scheduled the planet's ephemeris was still poorly known, so transit
began shortly after sunset; in addition, technical issues further
delayed the start of observations. A subset of instrumental parameters
during our observations is shown in Fig.~\ref{fig:obs}. Instrumental
seeing ($w$, measured by fitting a Gaussian profile along the spatial
direction of \gj's spectrum in each raw frame) varied from FWHM values
of 0.5--1.3''.  Spectral motions along the dispersion and spatial axes
were roughly 3~pixels each and mainly dominated by a gradual trend
attributable to differential atmospheric refraction owing to the use
of an optical-wavelength guider. A software upgrade under development
should correct for this differential refraction.

The stellar traces moved in the spatial direction ($y$) by about
3~pixels (excluding a brief 4-pixel jump and subsequent recovery)
during the observations. All of $w$, $x$ (movement in the dispersion
direction), and $y$ exhibit rapid changes near mid-transit, which
introduces a change in the measured fluxes which is not wholly removed
when dividing by the comparison star's flux.  This abrupt event
occurs close in time to a discontinuity in the secondary and telescope
focus positions. Telescope staff currently have no explanation for
this effect.  As we describe below we attempt to account for this
effect in our light curve fits using these parameters as decorrelation
variables. Future MOSFIRE observations of \gj\ may achieve better
stability by correcting for the refractive effects described above,
and by defocusing the telescope to decrease the sensitivity to
instrumental flexure and to allow longer integration times.

\subsubsection{Calibration}
We calibrate the MOSFIRE data using our own set of Python and PyRAF
routines\footnote{Most these routines are available from the primary
  author's website, currently at
  \url{http://www.mpia-hd.mpg.de/homes/ianc/}.}; many of the initial
steps are similar to those previously employed in calibration of
Subaru/MOIRCS imaging data \citep{crossfield:2012d}. These routines
construct two master flat frames by median-stacking individual frames
taken with 0.7'' slits while the dome lamps were on and off. The
difference of the two frames produces the master, unnormalized dome
flat frame.  To remove the gross signatures of instrumental throughput
and the lamp spectrum, we divide the spectral region from each slit by
a one-dimensional median-filtered spectral profile; we also
divide by the median spatial profile to remove the effects of
MOSFIRE's nonuniform slit widths as discussed in
Appendix~\ref{sec:mosfire}. The resulting master flat frame
approximately captures the pixel-to-pixel sensitivity variations in
the MOSFIRE detector.

Our calibration routines next divide each raw science frame by the
resulting flat frame. We then correct the frames for the infrared
array detector's intrinsic nonlinearity on a pixel-by-pixel basis,
using the approach of \cite{vacca:2004}; we compute the necessary
nonlinearity coefficients from a large set of flat frames generously
provided to us by Dr.~K. Kulas. Next we flag all isolated pixels that differ
from their neighbors by $\gtrsim10\times$ the deviations expected from
photon and read noise, and replace these presumably bad pixels with
the median of their four nearest good neighbors. The final step is to
subtract the thermal and sky background from each frame. For each
pixel in a given frame, we fit a linear trend in time to the same
pixel in the preceding and succeeding frames (which are taken at the
opposite nod position), interpolate to the time of the frame of
interest, and subtract the interpolated value.

Wavelength calibration is provided by a set of frames using the same
0.7'' slits used for the flat frames and taken with Ne and Ar arc
lamps illuminating the dome interior. We extract the line emission
spectra from the master arc frame, and use the standard IRAF task
\texttt{ecidentify} to mark known lines and fit low-order polynomials
of pixel number versus wavelength. We find that a cubic function
adequately fits the wavelength solution for all targets; the best-fit
coefficients vary somewhat across the detector but the mean solution
is $\lambda(x)/\micron = \lambda_0 + (2.1953 \times 10^{-4}) x + (2.68
\times 10^{-9}) x^2 + (7.6 \times 10^{-13}) x^3$, where $\lambda_0$ is
the wavelength offset for each star and $x$ is the pixel number. The
fit residuals have RMS values of roughly 0.1\,\AA, with maximum
excursions of 0.4\,\AA\ occurring near the ends of some spectra; the
effective size of the stars on the detector varies with seeing but is
approximately 13\,\AA. Considering the very wide spectral bandpasses
we employ below, this level of calibration is probably adequate for
our purposes. As a final step, we compare the wavelength-calibrated
spectra to a model telluric absorption spectrum to correct for any
offsets (as would be caused by slightly mis-centering a star in our
wide slits). The median MOSFIRE spectrum of \gj\ and our comparison
star are shown in Fig.~\ref{fig:rawspec}.

\subsection{Gemini-North/GMOS Optical Spectroscopy}
\subsubsection{Observations}
We observed \gj\ on UT 2013-03-29 with the GMOS multi-object optical
spectrograph \citep{hook:2004} at Gemini North. This instrument (and
its twin at Gemini South) has already been used for transmission
spectroscopy of several systems \citep{gibson:2013,stevenson:2013}. As
at Keck, the combination of uncertainty in \gjb's ephemeris and
technical problems caused our observations to begin just as transit
ingress started. We observed using integration times of 100~s, and
read out only a subset of the detector pixels to reduce overheads. In
this way we obtained 135~frames during 252~minutes and covering an
airmass range of 1.01--1.72. 

We observed in the red optical using the RG610 filter and R600
grating, which gives a spectral resolution of 3700 for a 0.5''
slit. Again, we used 10'' slits and TYC~1363-2087-1 as our sole
telluric calibrator; raw spectra of these stars are shown in
Fig.~\ref{fig:rawspec}.  Throughout the night we observed through thin
cirrus, which introduced relatively large-amplitude (but largely
wavelength-independent) variations in our final time series
data. \gjb's transit is clearly visible in the wavelength-integrated
relative light curve shown (along with other, generally stable,
instrumental trends) in Fig.~\ref{fig:obs}.

\subsubsection{Calibration}
To calibrate the raw GMOS frames we first bias-subtract them and  apply a
flat-field correction. The flat frames were taken with a mask
otherwise identical to our science mask, but with 0.7'' slits. For
wavelength calibration we acquired CuAr arc lamp frames with this same
mask, and we determine wavelength solutions in the same manner as for
MOSFIRE.

GMOS spectra are dispersed across three independent CCD detectors,
which exhibit slight misalignments. We therefore perform independent
wavelength calibration and spectral extraction for each detector. Our
final light curve analysis (described below) explicitly ignores data
taken near the detector gaps.  

\subsection{Comparison Stars}
\label{sec:comparison}
The object TYC~1363-2087-1 serves as our primary telluric calibrator
for both the infrared and optical analyses. This star has $K_s=8.05$
and lies 2.9' northeast of \gj; the latter's photometric colors are
rather redder. A comparison of the two stars' spectra show that
TYC~1363-2087-1 is a giant or subgiant with $T_\textrm{\tiny eff}$
greater than that of \gj. The comparison star exhibits stronger
absorption at Brackett~$\gamma$, in the CO bandheads at
$\lambda>2.29$\,\micron, and in the $\sim0.86\,\micron$ Ca~II triplet,
and it shows weaker absorption at the 0.82\,\micron\ and
2.206\,\micron\ Na~I doublets, Ca~I 2.263\,\micron\ triplet, and
red-optical TiO bandheads \citep{kleinmann:1986,rayner:2009}.

We also obtained infrared and optical spectroscopy of
2MASS~07591321+1524069; this red object's CO bandhead, Na, and Ca
absorption are all shallower than for \gj, so it may be a main
sequence star.  However, it is much fainter ($I=13.4, K_S=11.9$) than
our primary comparison star and so its light curves are too noisy to
be of further use in our analysis.

\subsection{Spectral Extraction and Final Calibration Steps}
We use the standard IRAF task \texttt{apall} to extract spectra from
our calibrated data frames and we explore a
wide range of extraction parameters: background and target aperture
sizes, and polynomial orders used for tracing the spectral traces and
for fitting residual sky background.  We employed this approach in two
parallel ways: in one the extraction aperture is constant for all data
frames, whereas in the other the extraction aperture for each frame
varies linearly with the spatial FWHM of the spectral trace in each
frame.  The optimal extraction parameters are those values which
minimize the RMS of the best-fit residuals in light curve fits to
several wavelength channels.

In our initial extraction parameter surveys, we find that the approach
with variable extraction aperture size typically gives slightly lower
residual RMS values. The parameters which minimize the residual
dispersion in our MOSFIRE analysis uses extraction apertures with a
total width of ($2\times\textrm{FWHM} + 16$) pixels, surrounded by an
8~pixel buffer and then a background aperture 45~pixels wide, traced
with a fifth-order polynomial and including a linear trend in the
background estimation and subtraction.  For the GMOS analysis we find
the best results using an extraction aperture 20~pixels wide, with a
6~pixel buffer on each side flanked by 24~pixels of background
aperture, tracing the spectra on each chip with a third-order
polynomial and assuming a constant sky background across the region of interest.

After extraction, the final step in all cases is to cross-correlate
each star's spectra with the median of all extracted spectra to
measure any spectral shift in the dispersion direction, and then shift
each spectrum to a common wavelength scale using sub-pixel
interpolation.  We convert each frame's timestamps in the raw FITS
file headers into the BJD$_{\textrm TDB}$ time system using the online
interface of \cite{eastman:2010}.


\section{Light Curve Analysis}
\label{sec:analysis}
Before describing our approach to fitting the wavelength-integrated
and spectroscopic light curves, we first discuss how we treat stellar
limb darkening (\S~\ref{sec:limbdarkening}), model fitting and
parameter uncertainty estimation (\S~\ref{sec:erranal}), model
selection (\S~\ref{sec:modelselection}), and the details of fitting
the wavelength-integrated (\S~\ref{sec:whitelight}) and spectroscopic
(\S~\ref{sec:speclight}) light curves. Finally, in
\S~\ref{sec:injection} we investigate systematic errors using a set of
injection and recovery tests.

\subsection{Limb Darkening}
\label{sec:limbdarkening}
A proper treatment of limb darkening is essential for an accurate
retrieval of light curve parameters. A common practice is to
use model stellar atmospheres to help constrain limb-darkening
parameters \citep[e.g.,][]{bean:2010,berta:2012}.  To this end, we
investigate  two approaches which give consistent results.

In both cases, we perform fits for each individual wavelength channels
of the model atmosphere; the results are averaged together after
weighting the fits by the mean observed spectrum of \gj.  In agreement
with previous studies \citep{diazcordoves:1992,vanhamme:1993}, we find
that a root-square darkening relation -- of the form $I(\mu)/I(1) = 1
- c(1-\mu) - d(1-\sqrt{\mu})$ -- represents the stellar models better
than a quadratic relation. Our data are not of sufficient precision to
justify the use of the full four-parameter functional form, while the
use of a linear law gives higher residual RMS values (though the
ultimate results of this latter analysis are consistent with our final
results). The limb-darkening priors used in our analysis are listed in
Table~\ref{tab:ldprior}.

Stellar limb-darkening coefficients were calculated by fitting
monochromatic intensities from a spherically symmetric PHOENIX stellar
atmosphere model \citep[following][]{hauschildt:1999} customized for
GJ~3470.  A stellar temperature of 3500~K, a surface gravity equal to
$10^5$~cm~s$^{-2}$, and solar abundances were adopted for the final
analysis.  These stellar parameters were motivated by previously
published values \citep{demory:2013} and a desire to closely match
optical to infrared broad-band photometry for this star.  We then fit
the root-square limb-darkening relation to the high-resolution output
of this model. This is the approach used for the results we present below.

By adjusting the stellar parameters ($\log_{10} g$, $T_\textrm{{\tiny
    eff}}$, and [Fe/H]) to generate several new models, then
re-fitting, we obtain estimates of how the uncertainties in the
stellar parameters influence our uncertainties in the limb-darkening
values.  We compute the covariance matrix of $c$ and $d$, and use this
with optimal model's parameters to impose a two-dimensional Gaussian
prior in the fitting process. This approach makes full use of our
insight into the expected correlations between multiple limb-darkening
parameters. Because different models can give different limb-darkening
profiles (especially for cooler stars), we multiply the covariance
matrix by 4 to account for systematic uncertainties in the stellar
atmosphere models.

We also investigated the use of Kurucz model atmospheres
\citep{kurucz:1979} for estimating limb-darkening coefficients.  We
draw a large number of values of $T_\textrm{\tiny eff}$ and $\log g$
from Gaussian distributions described by the parameters in
Table~\ref{tab:syspar}; we use the set of models with [Fe/H]=0.2,
appropriate for \gj's metallicity
\citep{demory:2013,pineda:2013,biddle:2013}. For each draw we
interpolate the four nearest models and compute the desired
limb-darkening coefficients. The mean and covariance matrix of the
distribution are then applied as described above. Using this second
set of models gives a transmission spectrum consistent with our
primary result.

\subsection{Light curve analysis: parameter fitting and statistical uncertainties}
\label{sec:erranal}
We model the light curves using the standard \cite{mandel:2002}
equations, with a root-square limb darkening law and the
``small-planet'' approximation (to speed computation time). We fit to
the data using a model of form
\begin{equation}
F(t) = f(t) p(t) + \sum_i k_i s_i(t),
\end{equation}
where $F(t)$ is the final model at time $t$, $f$ is the transit light
curve, $p$ is a polynomial function of $t$, and the $s_i$ are the
state vectors with coefficients $k_i$ used for additional
decorrelation (see \S~\ref{sec:modelselection}). Our model parameters
are the transit time $T_T$, $R_P/R_*$, $R_*/a$, $P$, $i$, $c$, $d$,
the $k_i$, and the coefficients of $p$ (variables not otherwise
defined here have their usual meaning).  In all our analyses, we
impose a Gaussian prior on the orbital period $P$, with mean and
standard deviation \porb\ \citep{biddle:2013}. Our final results are
insensitive to small changes in $P$; for example, our conclusions do
not change significantly if we instead take $P$ from the work of
\cite{demory:2013}.

Function optimization is accomplished using \texttt{optimize.fmin},
the SciPy implementation of the downhill simplex algorithm. After an
initial fit, data points whose absolute residuals exceed $10\sigma_d$
are flagged as outliers, where $\sigma_d$ is the width of the
symmetric interval enclosing 68.3\% of the data points. We then
perform a weighted fit in which the data points are given equal
weights, such that the total $\chi^2$ equals the number of valid
(i.e., non-outlier) points. Data points whose absolute residuals
exceed $5\sigma_d$ are set to zero weight and the process repeats;
typically only a single iteration is required and only a small number
of data points (3-4) are de-weighted.

We explore the likelihood distributions of our parameters of interest
using the affine-invariant Markov Chain Monte Carlo (MCMC) sampler
\texttt{emcee} \citep{foreman-mackey:2012}. This tool computes
parameter distributions using many (in our case, several hundred)
independent Markov Chains; its affine-invariance means that the
algorithm is unaffected by the parameter correlations that plague
transit light curve fitting. The software therefore requires little
fine-tuning.  We verify by eye that the resulting chains are well-mixed
by examining the evolution of their $\chi^2$ values. In each fit, we
take as our final parameter estimates the resulting best-fit
parameters from the optimizer.

In the highly multidimensional parameter spaces we explore, we find
that optimizers easily become stuck in local likelihood maxima. To
guard against this eventuality, after each 20\% segment of every MCMC
run we re-apply the function optimization described above using using
the most recent parameters from 12 of the chains. If any of the resulting fits
have a higher likelihood than our previous best fit, we update the
weights on the data points, re-initialize the MCMC sampler, and begin
the procedure all over again. This approach affords us some assurance
that the sampler and optimizer do not become trapped in local minima
in $\chi^2$ space.

In our experience, correlated noise is present in (at least) all
ground-based infrared observations of exoplanet transits and
occultations.  The $\chi^2$ statistic implicitly assume statistically
independent measurements, i.e. an absence of correlated effects, so
relying on this parameter (or the related BIC; see below) can result
in underestimated parameter uncertainties.  We employ three techniques
to investigate the level of any correlated noise in the data. The
first technique is the now-standard examination of how the residuals
to the best-fit model bin down over larger and larger time
intervals. We employ the $\beta$-statistic approach of
\cite{winn:2008} to inflate our initial uncertainties (estimated as
the lower and upper values corresponding to the 15.87\% and 84.13\%
percentiles of each parameters' distribution) based on the level of
correlated noise on the timescale of transit ingress.  To prevent the
result being influenced by small-scale structure, we compute $\langle
\beta_{T12} \rangle$, which we define as the mean of the $\beta$
values from half to twice the ingress timescale.  This approach
estimates our statistical uncertainties; we employ a separate
injection-and-recovery approach (described in \S~\ref{sec:injection})
to estimate our systematic uncertainties.  We eschew the use of the
common residual-permutation bootstrap (or prayer-bead) analysis. As
has been noted elsewhere, bootstrap techniques can
be poorly suited to sampling the entirety of a posterior distribution
\citep{line:2013}.

\begin{figure}[tb!]
\begin{minipage}{0.55\textwidth}
\centering
\includegraphics[width=4.75cm]{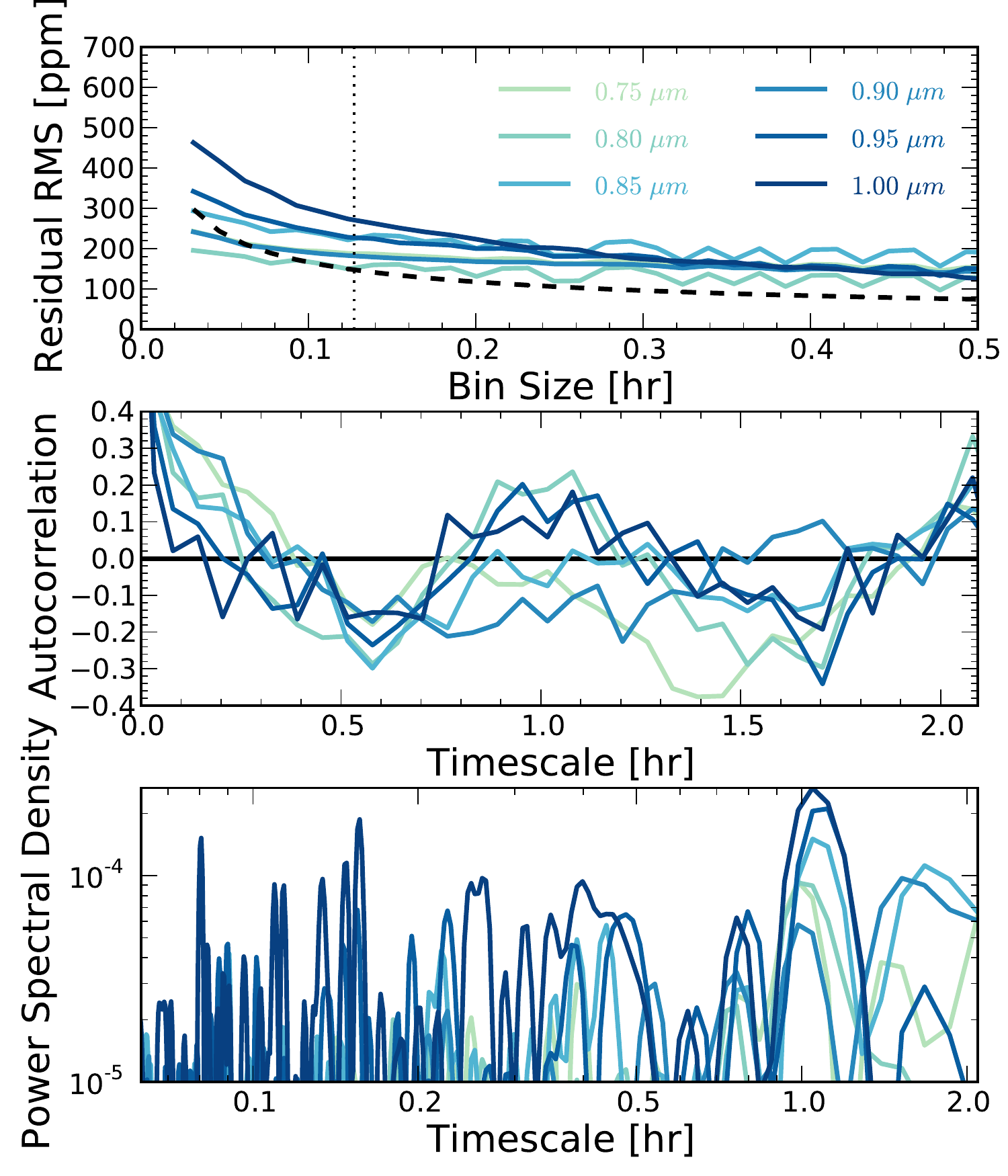}
\includegraphics[width=4.75cm]{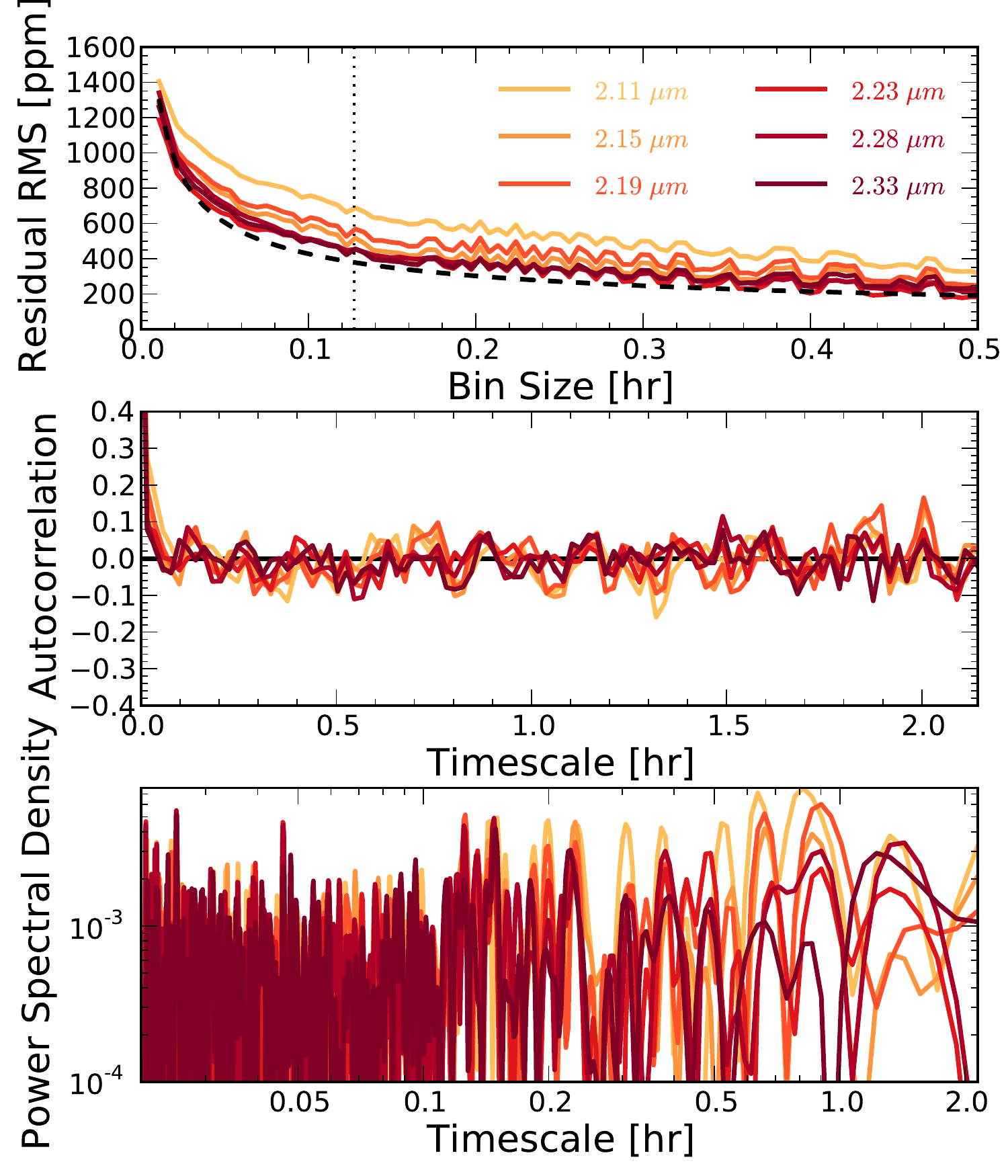}
\vspace{-0.2cm}
\end{minipage}
\caption{\footnotesize \label{fig:stats} Noise analysis of the residuals to the
  spectrophotometric light curves shown in
  Figs.~\ref{fig:mosfire_lightcurves} and~\ref{fig:gmos_lightcurves}
  for data from GMOS (left) and MOSFIRE (right). {\em Top:} Decrease
  in RMS of binned residuals with increasing binned size. The solid
  curves correspond to the individual wavelength channels, and the
  dashed line shows the 1/$\sqrt{t}$ expectation for independent and
  uncorrelated errors. The vertical dotted line indicates $T_{12}$,
  the duration of transit ingress. {\em Middle:} Discrete
  autocorrelation of the residuals. {\em Bottom:} Power spectral
  density of the residuals. The GMOS data exhibit strong internal
  correlations on 1-hour timescales, while the MOSFIRE data exhibit
  much weaker correlations.}
\end{figure}

Following \cite{gibson:2012}, we also always inspect the discrete
autocorrelation of the residuals and the power spectral density of the
residuals. Correlated noise induces large-scale structure in the
autocorrelation, and induces generally stronger peaks in the power
spectrum. These indications are more qualitative than quantitative,
but still give us greater confidence in our ability to {\reff
  ascertain whether any correlated noise is present}. We plot all of
these metrics for all spectrophotometric light curves in
Fig.~\ref{fig:stats}.

\subsection{Model Selection}
\label{sec:modelselection}
In the analysis of transit light curves, several techniques have been
used to address the issue of model selection, including Gaussian
Processes \citep{gibson:2012} and statistical Information Criterion
\citep{liddle:2007,stevenson:2010}. At least as often a single
parametric model is assumed, but there is no guarantee as to the
accuracy of model parameters and uncertainties so derived. We use the
Bayesian Information Criterion \citep[BIC=$\chi^2 + k \ln n$ when
fitting $n$ measurements with a $k$-dimensional
model;][]{schwarz:1978}, which penalizes models that use too many
parameters.

In our case, the extra parameters are the additional observables shown
in Fig.~\ref{fig:obs} that we use as the
$s_i$ in our analysis. We fit our model function to the white-light
(wavelength-integrated) light curve using many different combinations
of these state vectors and compute the BIC for each model. If one
naively follows the standard practice of rescaling our measurement
weights to give $\chi^2=n$, then a zero-parameter model will always
minimize the BIC. Instead, we follow our former approach
\citep{crossfield:2012d} and rescale our measurement weights based on
an initial, baseline fit. For both MOSFIRE and GMOS analyses, our
baseline model is the product of the standard light curve relation and
a quadratic function in time. We then compare the BIC resulting from
more complicated models to the BIC from this model.

\subsection{White light curve analysis}
\label{sec:whitelight}
We test various combinations of state vectors on our MOSFIRE data, and
in Table~\ref{tab:mosfirebic} list the resulting values in
$\chi^2$ and BIC, the $\langle \beta_{T12} \rangle$ values, and the
standard deviation of the baseline-normalized residuals (SDNR).  Our best
BIC-minimizing models listed in Table~\ref{tab:mosfirebic} all produce
results consistent within the reported uncertainties, so our precise
choice of model does not affect our ultimate conclusions.

Note that the use of BIC as a model selection tool is probably not the
optimal choice, because it ignores the correlated noise present in the
data. An ideal metric would penalize greater model complexity,
per-point residuals, and models whose residuals exhibit a greater
level of internal correlation.  A more rigorous approach to these
matters will eventually be necessary if the exquisitely precise
measurements planned with JWST are to be reliable
\citep[e.g.,][]{deming:2009,kaltenegger:2009}.

For the GMOS data, the scatter in the relative light curve (see
Fig.~\ref{fig:obs}) is much greater than would be expected from
photon noise considerations. This increased scatter is likely the
result of the thin cirrus layer present throughout the night, and to a
large degree it affects all wavelength channels equally. Because the
noise level is clearly very high (and the noise does not correlate
well with other observables), we adopt an essentially differential
approach for these data. For the wavelength-integrated analysis we use
a quadratic trend in time with no other state vectors. We then use the
residuals to this fit as the single state vector for analyzing the
spectroscopic light curves (as described in the next section).  Such
self-calibration can artificially reduce the noise level in each
wavelength channel. We therefore inflate all GMOS
spectroscopic-channel uncertainties by $\sqrt{1+1/v}$, where $v$ is
the number of spectroscopic channels used.

\subsection{Spectroscopic Analysis}
\label{sec:speclight}
After selecting the state vectors to use in decorrelating our
photometry ($x$ for MOSFIRE, and the white-light residuals for GMOS),
we perform an analysis of light curves constructed by splitting our
stellar spectra of target and comparison star into several bins. For
both optical and NIR analyses we choose to use six bins to strike a
balance between our final uncertainties and retaining some moderate
spectral resolution.

The spectroscopic analysis proceeds as follows.  First, in
succession we perform an initial fit to each  {\reff channel's light curve} using the
parameters from the white-light analysis as an initial guess (but
without using any state vectors). The fit residuals are again used to
set the per-point weights; points with zero weight in the white-light
analysis are also de-weighted in the per-channel analysis, and any
additional points more than $5\sigma$ discrepant are also de-weighted;
these steps repeat 1-2 times, until convergence.

Next, we compute a linear least-squares fit of the state vectors to
the residuals obtained. We combine the least-squares parameters with
the other parameters previously determined as a guess for a full fit
to the channel's light curve; again, we de-weight outliers and reset
the weights so that $\chi^2$ equals the number of data points. As
described previously, we then conduct an MCMC analysis of the channel
and interrupt the MCMC process several times during its run to test
for new and improved parameter sets. These steps are repeated for each
of the spectroscopic channels that are to be analyzed.

Now the real analysis begins: we analyze all spectroscopic light
curves in a single optimization and MCMC run. We use the weights
calculated previously for each channel, and use the best-fit
parameters from the individual fits as an initial guess.  A single
value controls each of the physical and achromatic parameters
($a/R_*$, $T_T$, $P$, and $i$) for all of the light curves, but $c$,
$d$, and $R_p/R_*$ are allowed to take different values in each
channel and the $s_i$ and coefficients of $p$ take on independent
values in each nod position.  With $N$ channels we have $7N+4$ free
parameters for the GMOS analysis and $11N+4$ for MOSFIRE. As before we
then fit this large number of parameters to the data, conduct an MCMC
analysis, and check several times for parameter combinations that
improve the quality of the fit. We then estimate statistical parameter
uncertainties as described in \S~\ref{sec:erranal}.

\subsection{Injection and Recovery Tests}
\label{sec:injection}
To best understand the accuracy and precision of our analysis, we
conduct a series of tests in which we attempt to recover signals of
known amplitude which we inject into the residual data.  Such injection
tests are commonly used to determine the performance of high-contrast
imaging programs \citep[e.g.,][]{nielsen:2010}, planet surveys using
radial velocity \citep[e.g.,][]{wittenmyer:2011} and transit
\citep[e.g.,][]{petigura:2013} methods, and exoplanet spectroscopy
\citep{crossfield:2011,crouzet:2012,deming:2013}.  We insert
artificial signals into the observed data, then re-analyze the
synthetic data sets using our standard techniques. Any differences
between the artificial signals' recovered parameters and the known
input values provide a quantitative measure of how our analysis
methods and the data's noise properties affect the parameters
retrieved from the non-injected analysis.

We implement the injection and recovery scheme as follows. First, we
take the extracted spectra and conduct the analysis described in the
preceding sections. We then remove the best-fit transit signature from
each wavelength channel (to provide sufficient coverage outside of
transit) while leaving all trends and systematic effects unaltered.
To the extent that removing the best-fit transit signature inevitably
reduces the noise level of the remaining data, the accuracy estimated
from our approach is overly optimistic; however, this is likely to be
a minor effect.

We next inject a simulated transit light curve computed by taking $P$,
$a/R_*$, $i$, and the limb darkening coefficients from the current
channel's best-fit parameters; $R_P/R_*$ and $T_T$ are each set to the
same values across all channels. $T_T$ in particular is chosen to be
offset by at least $T_{14}/2$ from its nominal location to avoid
re-fitting data sets which are essentially identical; in practice this
means our relatively short data sets limit us to $\sim$2 such
independent tests (i.e., signals injected halfway through the data and
at the end of the data). Because the signals are injected into the
absolute (not relative) spectrophotometry our approach also
automatically injects a signal into the white-light
(wavelength-integrated) light curve. We then run the analysis process
described in the preceding sections on the new, synthetic data set and
compare the results to the input parameters.  The primary difference
between our approach and residual permutation analysis is that we are
examining how the overall character of the measured transmission
spectrum is affected by systematic effects and correlated noise,
whereas residual permutation is typically used as an estimate of the
noise properties of an individual light curve, independent of other
wavelength channels.

We show the result of the MOSFIRE injection-and-recovery tests in
Fig.~\ref{fig:mosinj}. Both recovered spectra are flat (reduced
$\chi^2_{\textrm{\tiny flat}}<1$ when compared to a flat spectrum) and
show systematic offsets between the mean injected signal ($R_P/R_* =
0.07$) and the recovered values of $<2\sigma$. We note that the
recovered spectra both exhibit similar shapes, with extrema near
2.25--2.30\,\micron. Although this shape is correlated with the
limb-darkening priors applied in our analysis, we recover a spectrum
with the same shape even when imposing no limb-darkening prior. In
addition, \gjb's true recovered K-band transmission spectrum (shown in
Fig.~\ref{fig:specfig}) does not exhibit this shape. We estimate our
MOSFIRE systematic uncertainties to be $\sigma_{R_P/R_*}=0.03$
(absolute) and 0.16\% (relative), where the former is computed by
taking the weighted mean of the computed offsets and the latter by the
root quadrature mean of the RMS from the recovered spectra; both
values are lower than the per-channel uncertainties presented in
Table~\ref{tab:specdat}. All these encouraging factors lead us to
conclude that at our current level of precision, our MOSFIRE spectrum
is not significantly affected by systematic sources of error. However,
future observations with better pre- and post-transit coverage should
reach higher precision and these tests should be repeated then.

The GMOS injection-test results are shown in
Fig.~\ref{fig:gmosinj}. As with MOSFIRE, the injected signal has
$R_P/R_*=0.070$. When the signal is injected into the middle of the
data set, the resulting spectrum is quite flat
($\sigma_{R_p/R_*}=0.13\%$ and reduced $\chi^2_{\textrm{\tiny
    flat}}<1$). However, when we inject a signal at the end of the
observations, the recovered spectrum is worse
($\sigma_{R_p/R_*}=0.16\%$, reduced $\chi_{\textrm{\tiny
    flat}}^2=1.652$). With so few wavelength bins this
$\chi_{\textrm{\tiny flat}}^2$ value could occur by chance in 14\,\%
of cases, but it is still troubling.  The systematic offsets between
the injected signal and the mean recovered value are quite large,
confirming our hypothesis that the differential analysis used here is
not sensitive to the absolute transit depth. We estimate systematic
uncertainties of 0.75\,\% (absolute) and 0.14\,\% (relative) for the
GMOS data.  Most troubling, the shape and amplitude of the spectrum
recovered from the observed (non-injected) GMOS data is quite similar
to that of the spectra recovered from the injection tests: a slope
below 0.90\,\micron\ with a local extremum near that wavelength.  We
therefore conclude that the GMOS data is likely to be contaminated by
systematic errors induced by the variable cloud conditions that
prevailed when during these observations.

\subsection{Transit Analysis: Results}
\label{sec:results}
\subsubsection{MOSFIRE K Band Analysis}
We restrict the range of our MOSFIRE data to 2.09--2.36\,\micron\
because data outside this range show increasingly strong systematic
errors in the spectroscopic analysis; this effect could be due to a
slight mismatch in wavelength calibration and correlations with
telescope pointing and instrumental seeing.

Using this bandpass, the results from our MOSFIRE white-light
(wavelength-integrated) are listed in Table~\ref{tab:syspar}. Our
best-fit orbital parameters ($i$ and $a/R_*$) are consistent with
those presented previously \citep{demory:2013,fukui:2013} and we find
$R_p/R_* = \rprs$, which is $<1.5\sigma$ discrepant from all
previously reported values for \gjb.

\subsubsection{Retrieved Transmission Spectrum}
The MOSFIRE light curves, best-fit models, and residuals to the fits
are shown in Fig.~\ref{fig:mosfire_lightcurves} and the model
parameters are listed in Table~\ref{tab:specdat}. The residual scatter
in these fits is 2--3$\times$ greater than predicted by Poisson
statistics. We find that we come closer to the photon noise limit in
regions free of telluric OH line emission, consistent with previous
analyses \citep{bean:2011}.

\fig{mosfire_lightcurves}{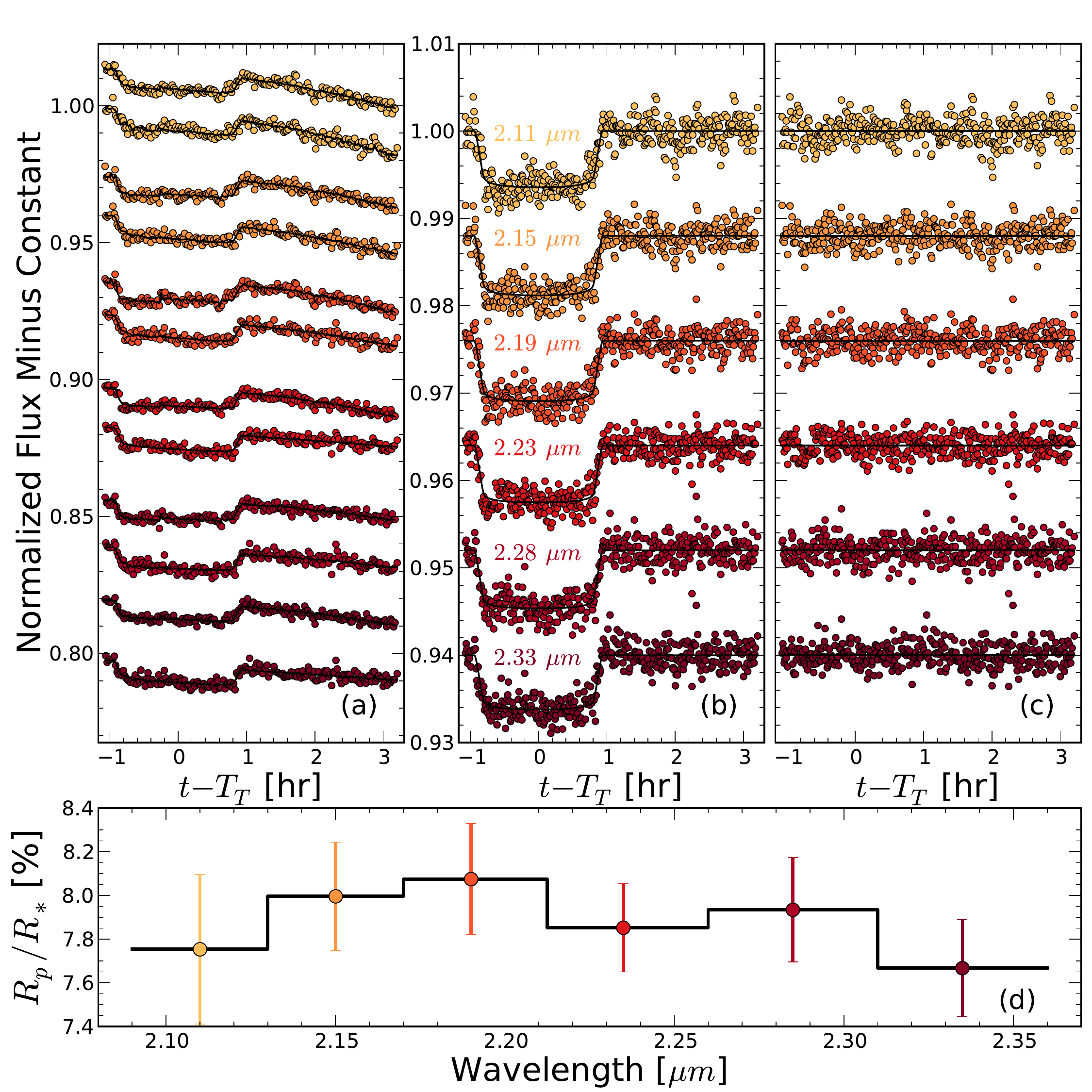}{width=9cm}{}{\footnotesize :\vspace{0.05in}\\MOSFIRE
  transit light curves: (a) raw spectrophotometric data showing the
  systematic offsets between the two nod positions, and best-fit
  models; (b) normalized and corrected measurements and models; (c)
  residuals; (d) final transmission spectrum after scaling error bars
  by $\langle \beta_{T12} \rangle$. }

Because GMOS uses three CCDs, there are slight gaps (corresponding to
light lost between the CCDs) in our optical analysis; we split each
detector into two, giving six wavelength channels. Because of our
differential (self-calibrating) approach our GMOS transit depths are
precisely determined relative to the determined white-light transit
depth, but their absolute accuracy is limited to the accuracy of the
low S/N parameters from the noisy wavelength-integrated light
curve. The GMOS light curves, best-fit models, and residuals to the
fits are shown in Fig.~\ref{fig:gmos_lightcurves}. As discussed in
\S~\ref{sec:injection}, the GMOS observations were contaminated by
cirrus-induced systematic effects and we do not consider these data
further.

\fig{gmos_lightcurves}{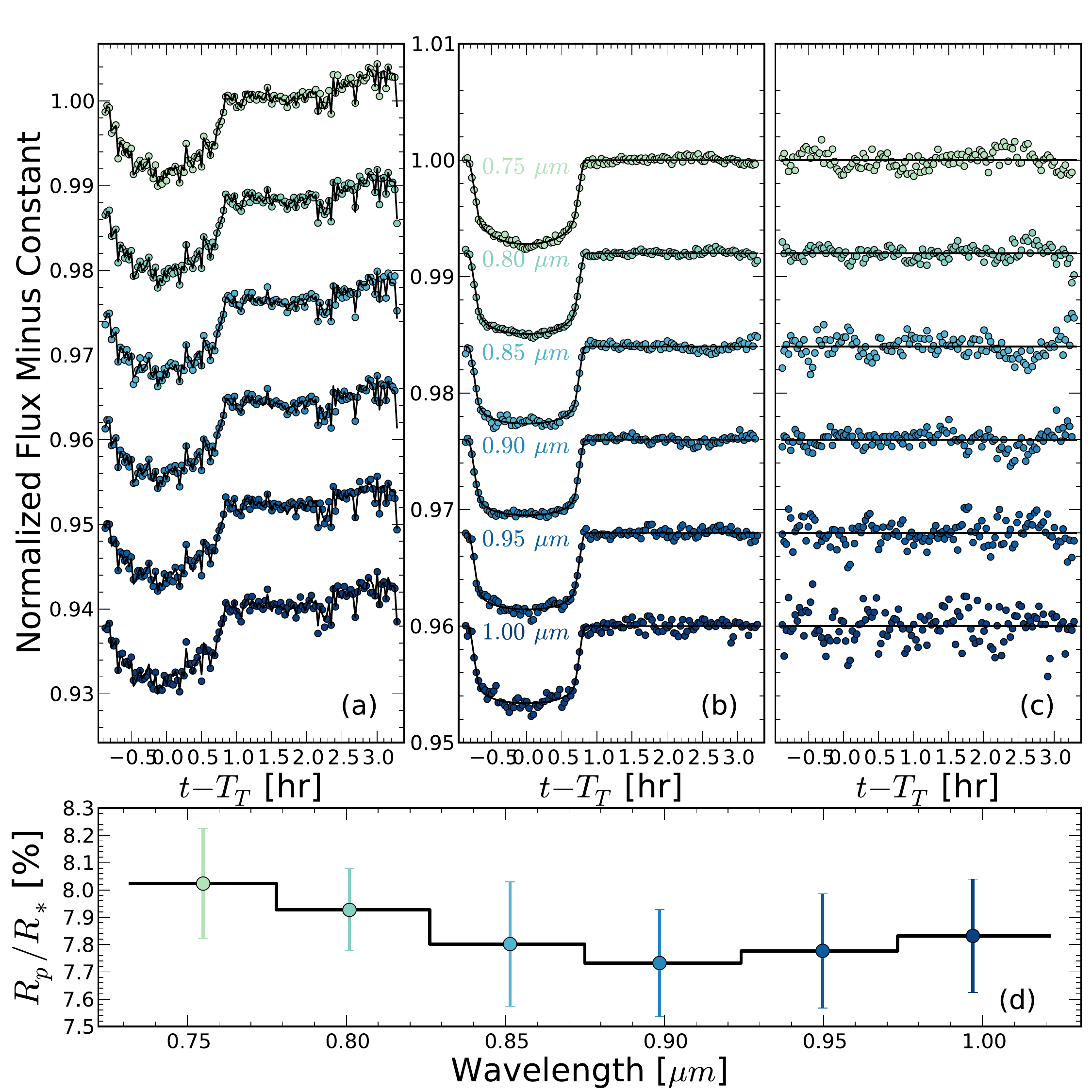}{width=9cm}{}{\footnotesize :\vspace{0.05in}\\ GMOS
  transit light curves: (a) raw spectrophotometric data showing the
  common-mode flux variations, and best-fit models; (b) normalized and
  corrected measurements and models; (c) residuals multiplied by 3;
  (d) final transmission spectrum after scaling error bars by $\langle
  \beta_{T12} \rangle$. Because we analyze the GMOS data in a
  differential fashion (\S~\ref{sec:whitelight}) these data are
  insensitive to the absolute transit depth. Further, as shown in
  Fig.~\ref{fig:gmosinj} these measurements are likely affected by
  systematic errors.}

\fig{specfig}{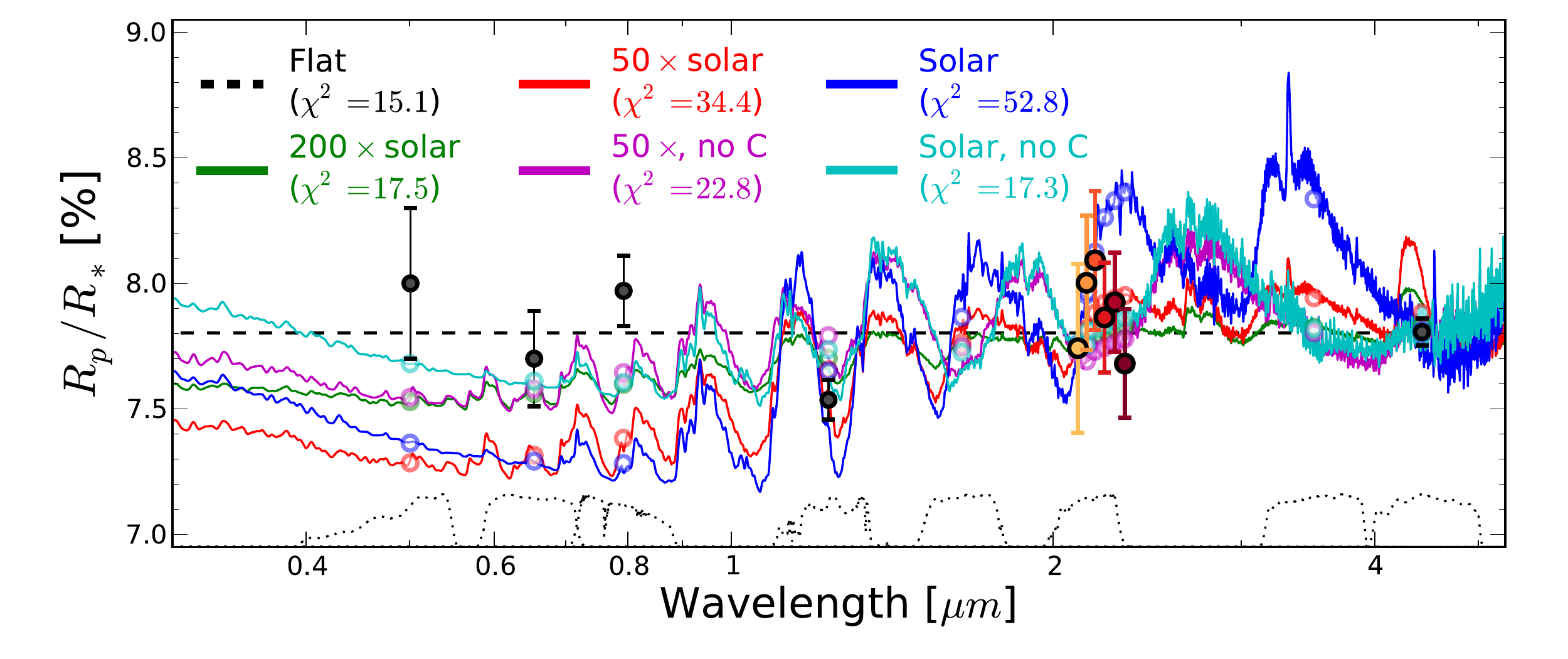}{width=9.5cm}{}{\footnotesize :\vspace{0.1in}\\Transmission spectrum of
  \gjb. Colored points with error bars are our MOSFIRE measurements;
  black points are the measurements of \cite{fukui:2013} and
  \cite{demory:2013}. The solid lines show the model transmission
  spectra described in \S~\ref{sec:disc}. The ensemble of measurements
  rule out equilibrium-chemistry models with solar composition (blue)
  and 50$\times$ solar abundances (green) at 5.4$\sigma$ and
  3.8$\sigma$, respectively.  A methane-depleted atmosphere (pink), a
  very highly enriched atmosphere (200$\times$ solar, light blue), and
  a simple flat spectrum suggestive of high-altitude haze (dashed) all
  remain plausible. These scenarios are discussed in
  \S~\ref{sec:disc}. The dotted lines at bottom show the filter
  profiles (from \cite{fukui:2013}, 2MASS, and Spitzer/IRAC) used to
  compute the band-integrated model points (shown as colored open
  circles).  }

\section{Atmospheric Models and Observations}
\label{sec:model}
\subsection{Model Parameters and Spectra}
\label{sec:atmoparam}
To interpret our observations, we present the first model transmission
spectra calculated specifically for \gjb. These models include
cloud-free PHOENIX models in chemical equilibrium with $1\times$,
$50\times$, and $200\times$ solar abundances, and an otherwise
identical set of models with the abundance of carbon set to zero
\citep[to simulate an atmosphere in severe chemical disequilibrium or
with a low C/O ratio, as has been investigated for
GJ~436b;][]{line:2011,madhusudhan:2011c}. We motivate our choice of
200$\times$ solar as an upper limit in \S~\ref{sec:himmw}. All models
are calculated following the steps outlined by \cite{barman:2007},
with the only exceptions being the use of more recent CH$_4$ line data
\citep{sromovsky:2012,bailey:2011} and new collision-induced opacities
\citep{richard:2012}.  We plot the temperature-pressure profiles and
vertical abundance profiles of several molecular species in
Fig.~\ref{fig:profiles}. Note that these models do not include the
effect of photochemistry or disequilibrium processes, which can
significantly affect atmospheric abundances in planets orbiting M
stars \citep[e.g.,][]{line:2011,moses:2013}.

\fig{profiles}{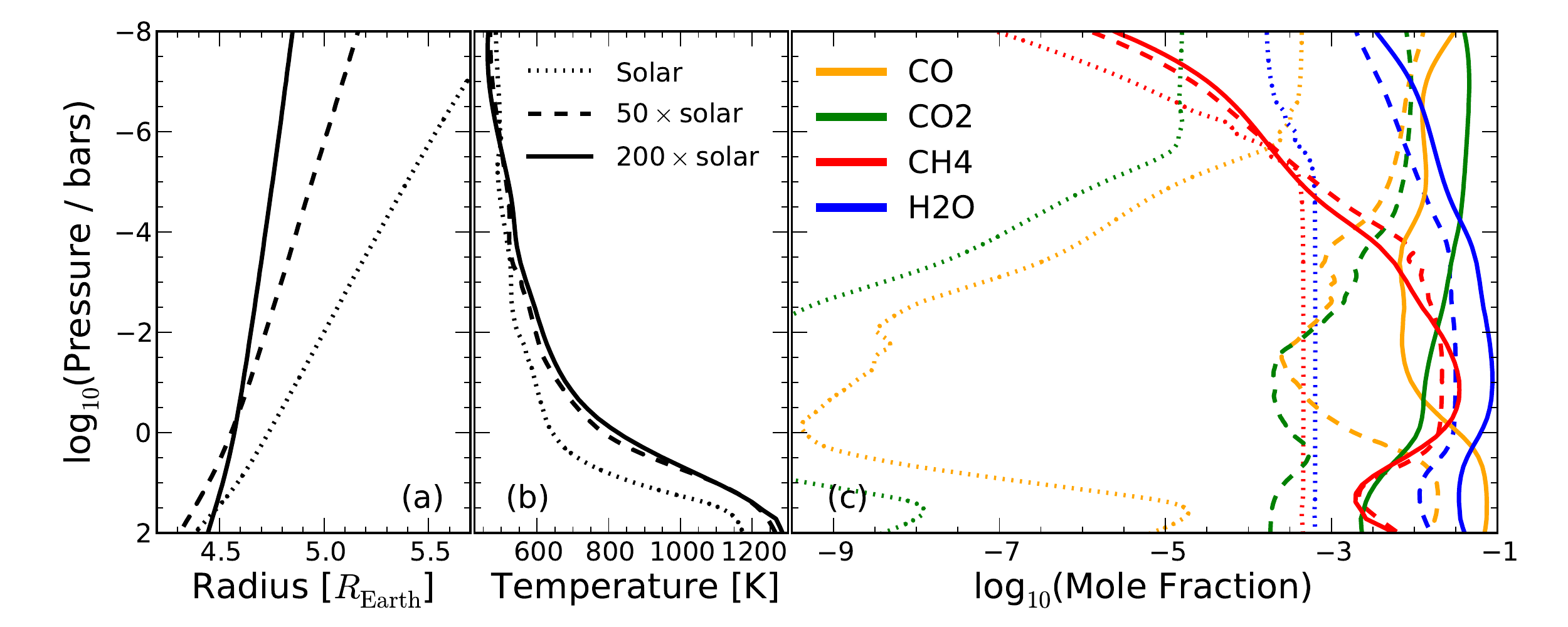}{width=9.5cm}{}{\footnotesize :\vspace{0.05in}\\ Atmospheric parameters for
  our equilibrium-chemistry models assuming solar abundances of heavy
  elements (solid), 50$\times$ solar (dashed), and 200$\times$ solar
  (dotted).  The panels show as a function of altitude: (a) the
  effective planetary radii, (b) the temperature-pressure profiles,
  and (c) the abundances of several molecular species.  We discuss
  these models in \S~\ref{sec:model}, and show the model transmission
  spectra derived from them in Fig.~\ref{fig:specfig}.  }

The equilibrium chemistry changes significantly across the metallicity
range considered.  With solar abundances ratios CH$_4$ is the most
abundance C-bearing molecule by far. As the abundances of heavy
elements are increased, the abundances of O-bearing species increases
more quickly than that of CH$_4$ (because the solar C/O ratio
$<1$). Thus in the 200$\times$ solar model, CO$_2$ is more common than
both CO and CH$_4$ throughout most of the atmosphere. H$_2$O remains
the most abundant molecule (after H$_2$) in all models at pressures
$\lesssim 0.1$~mbar. The main effects of increased metallicity in the
model spectra, shown in Fig.~\ref{fig:specfig}, are therefore: a
reduction in CH$_4$ features throughout the infrared spectrum; the
steadily strengthening CO$_2$ opacity feature at 4.3\,\micron; lower
amplitudes for features throughout the spectrum owing to the greater
mean molecular weight and smaller scale height.  The C-free models
have H$_2$O mole fractions of roughly $6.4\times10^{-4}, 0.031$, and
0.12, and these values are very nearly constant with altitude.

\subsection{Comparing Models to Data}
Fig.~\ref{fig:specfig} shows these model transmission spectra and the
observations of \gjb.  The figure also presents the predicted transit
depths in several photometric bandpasses (computed by weighting the
models with the stellar flux incident in each bandpass).

With our new MOSFIRE measurements, transit observations of GJ~3470b
currently comprise eleven measurements: our six spectroscopic points
and the previous five photometric measurements \citep[][we neglect the
discovery transit measurements of \citeauthor{bonfils:2012}~2013 owing
to the poor quality of those data]{fukui:2013,demory:2013}. We cannot
discriminate between our models on the basis of the MOSFIRE data
alone, but using all available data we compute the $\chi^2$ statistic
for each of the atmospheric models presented above. We tune each model
only inasmuch as we add a constant value to match the weighted mean of
all measurements. Table~\ref{tab:models} shows the resulting $\chi^2$
values and the probability $P(\chi^2)$ of each $\chi^2$ value occurring
by chance.

Of all models tested, the lowest $\chi^2$ results from a planetary
radius that is constant with wavelength: i.e., a flat transmission
spectrum. Such a model gives $\chi^2=15.1$, indicating that it is
reasonably consistent with the data. The C-free models have
$\chi^2\lesssim20$; worse than the flat case, but close enough that we
cannot distinguish between these cases.

Finally, the two lower-metallicity models provide the worst fits: the
50$\times$ solar abundance and solar models give $\chi^2=34.4$
and~52.8. Thus, the data disfavor these models at confidence levels of
3.8$\sigma$ and 5.4$\sigma$, respectively, when considering the
ensemble of all data.

\section{Interpreting A Flat Transmission Spectrum}
\label{sec:disc} 
 
The transmission spectrum of GJ~3470b is flat within current
measurement uncertainties.  Such a result admits of several
interpretations. Below we discuss several possibilities.  The planet
could have no (or a negligible) atmosphere (\S~\ref{sec:noatmo}); some
or all of the observations to date could suffer from undetected
systematic errors (\S~\ref{sec:measerr}); or the planet's atmosphere
could be: dominated by molecules heavier than H$_2$
(\S~\ref{sec:himmw}); hydrogen-dominated but with less methane than
expected (\S~\ref{sec:lomethane}); or enshrouded in optically thick
clouds or haze (\S~\ref{sec:haze}).

\subsection{No or Negligible Atmosphere?}
\label{sec:noatmo}
If \gjb\ had only a tenuous atmosphere, or no atmosphere whatsoever,
its radius would naturally appear nearly constant across the
wavelengths probed. However, models of planetary interiors predict
that a planet with \gjb's observed mass and radius should consist of
$\gtrsim10\,\%$ hydrogen by mass
\citep{fortney:2007,adams:2008,figueira:2009,valencia:2011}; whether
primordial or outgassed, such a massive atmosphere clearly indicates
that a substantial planetary atmosphere is preferred for this planet.

\subsection{Measurement Errors?}
\label{sec:measerr}
Early transit photometry of GJ~1214b showed a transit significantly
deeper in K band than at optical and Spitzer/IRAC wavelengths, a
result which was interpreted as evidence for a H$_2$-dominated
atmosphere depleted in CH$_4$ and/or with a low carbon content
\citep{croll:2011b,crossfield:2011}; a new, independent K band
photometric data set is consistent with this early result
\citep{demooij:2012}. However, K band transit spectroscopy of GJ~1214b
shows a transit shallower than the photometrically-derived values
\citep{bean:2011}.  The discrepancy has not yet been resolved, despite
the significant implications for the composition of GJ~1214b's
atmosphere.  The controversy highlights the need for multiple and
independent transit analyses when retrieval of atmospheric parameters
is desired.  In a similar manner, future observations of \gjb\ may
reveal a discrepancy in one or more of the transit observations we
consider here.

We verified that a flat transmission still fits better than other
models, even when we impose no priors whatsoever on the stellar limb
darkening.  Our results change only slightly if we perturb the prior
imposed on $P$, or if we impose priors on $i$ and $a/R_*$ using
results from previous analyses \citep{demory:2013}. Thus, the MOSFIRE
transit measurements appear robust.

Our own optical-wavelength transit light curve analysis
\citep{biddle:2013} gives a transit depth consistent with previous
measurements \citep{fukui:2013}. \gj's 1\,\% R band variability
\citep{biddle:2013} should not induce significant systematic offsets
between transits measured at different epochs, considering the
measurement uncertainties available.  We note that previous analyses
of GJ~3470b's transit light curves {\reff did not report the magnitude
  of correlated noise on their measurement uncertainties
  \citep{demory:2013,fukui:2013}}, so additional analyses and
observations clearly warranted\footnote{\reff B.-O. Demory informs us
  that their analysis estimated that correlated noise was a $<10\%$
  effect in their data using the $\beta$ approach of
  \cite{winn:2008}. }. At present, we see no cause to mistrust the
data in hand.

\subsection{An Atmosphere With High Mean Molecular Weight}
\label{sec:himmw}
Atmospheres which contain a large percentage of molecules heavier than
H$_2$ will have smaller scale heights, an effect which reduces the
amplitude of spectral features seen in transmission
\citep[e.g.,][]{miller-ricci:2009}. For example, GJ~1214b's
mostly-flat transmission spectrum has been interpreted as evidence for
an atmosphere with a large H$_2$O component
\citep[][]{bean:2011,berta:2012,howe:2012,benneke:2013}.  Our
200$\times$ solar abundance model shows spectral features somewhat
smaller in amplitude than those of our solar-abundance model
(Fig.~\ref{fig:specfig}) because of the former's higher mean molecular
weight; this high-metallicity model is consistent with the existing
data.

It has been suggested that the formation of low-mass, low-density
planets by core accretion could lead to extremely high atmospheric
enrichment in heavier elements \citep[$>100\times$
solar;][]{fortney:2013}.  \gj's supersolar metallicity
\citep{demory:2013,pineda:2013,biddle:2013} lends further support to
the idea that \gjb's atmosphere could be metal-rich\footnote{Following
  the astronomical convention of naming elements heavier than helium,
  ``metals.''}, perhaps more so than those of solar system ice giants.

In particular, a variety of theoretical models predict that planets
with \gjb's mass would form with high envelope
metallicities. Population synthesis models of planet migration predict
envelope metallicities of $Z = 0.6 - 0.9$, corresponding to
atmospheric mean molecular weights of roughly 5--12 \citep[assuming
H$_2$O is the dominant heavy molecule;][]{fortney:2013}. Models of in
situ accretion \citep{hansen:2012} predict global metallicities of
$Z=0.3-0.6$, prior to envelope evaporation.  However, the bulk
properties of \gjb\ constrain the metallicity of the atmosphere to be
not much above $300\times$ solar abundances ($\mu = 9$). We arrive at
this estimate by assuming: GJ 3470b's bulk composition is 10\% H/He by
mass \citep{demory:2013}, the planet's heavy-element content is composed
solely of H$_2$O, any metals not in the core are distributed evenly
throughout the atmosphere, and the planet's radius does not change
when metals are moved from the core to the envelope. These assumptions
may not all hold, but the point remains that GJ 3470b's low bulk
density sets an upper limit to the planet's total metal content. If we
further assume that the planet formed via gas accretion onto a solid
core of roughly 5 $M_\Earth$, the planet's atmospheric abundances must
be $\lesssim200 \times$ the solar level.

As Table~\ref{tab:models} and Fig.~\ref{fig:specfig} show, such a high
enrichment is consistent with the data.  Even higher metallicities (up
to $10000\times$ solar) have been recently proposed to explain
observations of GJ~436b's atmosphere \citep{moses:2013}. If low-mass
planets like \gjb\ and GJ~436b can indeed form with such highly
metal-enriched atmospheres \citep{fortney:2013}, then \gjb's lower
density and consequent tighter constraints on envelope metallicity may
make this system a more attractive target for transmission
spectroscopy than GJ~436b.  For now, a metal-enriched
{\reff ($\sim$200--300$\times$ solar}), high-mean molecular weight atmosphere
seems a plausible explanation for GJ~3470b.

\subsection{Methane-poor atmosphere}
\label{sec:lomethane}
Methane is a strong opacity source at 2.1--2.4\,\micron\ and is
predicted to be the dominant C-bearing molecule at temperatures of
600--800~K. In chemical equilibrium, a solar composition atmospheres
at these temperatures has a CH$_4$ mixing ratio of $>10^{-4}$ for
pressures $P=1-1000$~mbar and $>10^{-3}$ for a $30\times$ solar
metallicity atmosphere. At all but the upper
end of this temperature range, CO is predicted to be less abundant
than CH$_4$. Indeed, our lower-metallicity models of GJ~3470b's
atmosphere predict CH$_4$ to be more dominant than CO at pressures
$<0.01$~mbar (Fig.~\ref{fig:profiles}).

However, we see no evidence for the large differential ([4.5] -- $K$)
transit depth expected for an atmosphere with strong CH$_4$ absorption
\citep[e.g.,][]{miller-ricci:2010,crossfield:2011}. This indicates
that CH$_4$ opacity does not contribute significantly to the
transmission spectrum.  Such a scenario can be explained by
intrinsically low C abundances \citep{madhusudhan:2012b}, by
disequilibrium processes such as photochemistry and eddy diffusion, or
by a highly metal-enriched atmosphere (which will tend to favor CO and
CO$_2$ over CH$_4$, as shown in Figs.~\ref{fig:profiles} and
Fig.~\ref{fig:specfig}).

The situation appears similar to that of GJ~436b, which is only
slightly warmer than GJ~3470b and also shows no evidence for the
predicted CH$_4$ absorption in transmission or emission on the basis
of Spitzer photometry at $>3\,\micron$
\citep{stevenson:2010,knutson:2011}.  These observations of GJ~436b
have been attributed to a high-metallicity ($\gtrsim10\times$ solar)
atmosphere with internal diffusion and perhaps photochemistry also
playing a role \citep{line:2011,madhusudhan:2011c,moses:2013}. Strong
parallels also exist between these atmospheric properties and those of
Uranus and Neptune \citep{lunine:1993}, though to date models cannot
explain the extreme level of disequilibrium chemistry inferred in
GJ~436b's atmosphere.  Nonetheless, some combination of low C/O
and/or chemical disequilibrium by a combination of convection and/or
diffusion \&\ photochemistry could explain GJ~3470b's current
transmission spectrum.

\subsection{Hazy or Cloud-covered Atmosphere}
\label{sec:haze}
The best-fitting model transmission spectrum is a constant value from
0.5--5.0\,\micron, indicating that currently no spectral features are
detected in GJ~3470b's atmosphere. A flat transmission spectrum is
also a simple approximation for an atmosphere largely or partially
obscured by optically thick clouds or hazes
\citep[e.g.,][]{crossfield:2011,berta:2012}.  The presence of
significant hazes and/or clouds has been predicted to be a ubiquitous
feature of externally irradiated exoplanet atmospheres
\citep{fortney:2005,fortney:2013}. Indeed, such atmospheres have been
invoked to explain observations of cool, low-mass planets such as
GJ~1214b \citep{howe:2012,morley:2013} and of hotter, more massive
planets such as hot Jupiter HD~189733b \citep{pont:2013}. The
hypothesis is that either condensate clouds \citep[analogous to those
found in brown dwarf atmospheres; e.g.,][]{woitke:2004,freytag:2010}
or photochemical hazes \citep[similar to those found in Solar System
gas giant atmospheres; e.g.,][]{pilcher:1977,pollack:1987} could form
stable layers at high altitudes where they would significantly alter
the character of the planet's transmission and emission spectra.

Condensate clouds have received considerable attention owing to their
importance in brown dwarf atmospheres, but hazes in cooler
($<1000$~K), externally-irradiated exoplanet atmospheres have not
undergone much study.  However, the recent excellent study undertaken
by \cite{morley:2013} indicates that physically motivated models can
explain the approximately flat transmission spectrum of GJ~1214b:
either clouds in a high-metallicity atmosphere or hydrocarbon haze
composed of sub-\micron\ sized particles. Our model atmosphere
parameters shown in Fig.~\ref{fig:profiles} are only slightly cooler
than those used by \cite{morley:2013}, suggesting that their results
are applicable to GJ~3470b. This similarity indicates that GJ~3470b's
flat transmission spectrum could be explained by either condensate
clouds or hazes produced by hydrocarbon photolysis.

\section{Conclusions and Future Work}
\label{sec:conclusion}
\subsection{Conclusions}
Our observations provide the best constraints to date on the
atmosphere of the cool, sub-Neptune mass ice giant GJ~3470b.  The high
S/N possible for GJ~3470b will lead to this planet becoming a
touchstone object that will strongly influence our understanding of
cool, low-mass planetary atmospheres. The transmission spectroscopy
presented here represents the first step toward that goal. Our K band
spectroscopy, combined with optical and Spitzer photometry, allows us
to rule out a solar-abundance atmosphere in chemical equilibrium and
without clouds or hazes with 5.4$\sigma$ confidence
(Fig.~\ref{fig:specfig}). A similar model with 50$\times$ solar
abundances is also disfavored, albeit with  lower confidence
(3.8$\sigma$).

However, the precise nature of GJ~3470b's atmosphere remains uncertain
because the ensemble of measurements are reasonably well fit by a
single value of $R_p/R_*$, independent of wavelength; i.e., the
transmission spectrum is flat. After considering many possible
explanations (Sec.~\ref{sec:disc}), we conclude that GJ~3470b's
atmosphere is either depleted in CH$_4$ relative to equilibrium
expectations and solar abundances (as has been previously claimed for
the hotter and more massive GJ~436b; \S~\ref{sec:lomethane}), is
enshrouded in an optically thick cloud or haze layer (as suggested for
the smaller, cooler GJ~1214b; \S~\ref{sec:haze}), or is highly
enriched in metals (as suggested for both GJ~436b and GJ~1214b;
\S~\ref{sec:himmw}).  We note that these explanations are not mutually
exclusive: for example, the atmosphere could have a high metallicity
and also host clouds or haze, while a metal-rich atmosphere would
naturally result in low levels of CH$_4$.

We have presented the first exoplanetary results obtained with the new
MOSFIRE instrument, and present detailed discussion of some of 
MOSFIRE's idiosyncrasies (Appendix~\ref{sec:mosfire}). The
precision of our MOSFIRE data are limited by having little pre-transit
coverage, but our injection and recovery tests
(\S~\ref{sec:injection}) indicate that our MOSFIRE results are not
significantly affected by systematic  sources of error.
Intermittent cirrus interferes with our optical spectroscopy
observations, and our analysis indicates that these measurements are
probably contaminated by cloud-induced systematic errors. This
indicates that the self-calibration technique adopted in several
analyses \citep{gibson:2013,stevenson:2013} gives inconsistent results
in the limit of very large systematic variations. Those analyses did
not attempt transit injection and recovery tests, but we strongly urge
future researchers to do so in order to empirically and convincingly
determine the amplitude of systematic errors.

\subsection{Future Work}
Further observations of GJ~3470b at optical and infrared wavelengths
are clearly warranted. Further transit photometry with Warm Spitzer
and from ground-based facilities will be of great use in refining the
planet's orbital and bulk physical properties
\citep[][]{biddle:2013}. Such observations might help to characterize
the planet's broadband transmission spectrum: e.g., 3.6\,\micron\
observations could provide an independent confirmation of CH$_4$
depletion and sufficiently precise 4.5\,\micron\ observations could
constrain the high abundances of CO$_2$ predicted for high-metallicity
atmospheres (see Fig.~\ref{fig:specfig}).  Caution is warranted in
light of troubling inconsistencies in similar transit photometry of
GJ~436b \citep{beaulieu:2010,knutson:2011}, and it remains to be seen
whether GJ~3470's stellar variability will permit a meaningful
comparison of transit photometry taken at different epochs.

In this most pessimistic case, transit spectroscopy and simultaneous
multiband photometry would be the only way to reveal the nature of
GJ~3470b's atmosphere. Additional H and K band transit spectroscopy
would confirm our nondetection of CH$_4$. Transmission spectroscopy at
0.5--0.8\,\micron\ and at J and Y bands seems best suited to
discriminate between the different plausible atmospheric scenarios
(see Fig.~\ref{fig:specfig}). Optical observations could provide
strong constraints on the presence of any optically thick clouds or
hazes; in the absence of such phenomena, the absorption signatures of
H$_2$O could be descried in the red optical and shorter-wavelength
observations could measure the atmosphere's Rayleigh slope and thereby
independently constrain the mean molecular weight of the planet's
atmosphere.

Detection of clouds or haze could suggest a high albedo that would
decrease the received stellar energy input and increase the level of
internal heating necessary to explain the planet's luminosity.
Further RV observations, or ideally occultation timing measurements,
would place better limits on the level of any tidal heating.
Currently, detection of GJ~3470b's occultations is only feasible with
Warm Spitzer; such measurements would allow a direct comparison with
measurements of GJ~436b's dayside emission and would provide a strong,
independent test of the relative abundances of CH$_4$ and CO in
GJ~3470b's atmosphere.  The degree of similarity between GJ 436b's and
GJ~3470b's atmospheres, both in terms of carbon-species chemistry
and/or optical hazes, will be an important benchmark test for the next
generation of atmospheric models.

\section*{Acknowledgements}
We thank P.~Cubillos and Dr. J.~Harrington for in-depth discussions
about the finer points of light-curve fitting and systematic
uncertainties, Dr. H.~Knutson for discussions of GJ~436b and its
transmission spectrum, Dr. M.~Kassis, J.~Aycock, Dr. P.~Hirst, and
B.~Walp for their assistance in obtaining our MOSFIRE and GMOS
observations, and Drs. I.~McLean and K.~Kulas for useful discussions
regarding the MOSFIRE instrument.

This material is based upon work supported by NASA Origins of Solar
Systems under Grant No. NNX10AH31G awarded to TB.


Facilities used: Keck/MOSFIRE, Gemini/GMOS.

\clearpage

\begin{deluxetable}{c c c c}
\tabletypesize{\scriptsize}
\tablecaption{  \gjb\ Transit Parameters (2.09--2.36\,\micron) \label{tab:syspar}}
\tablewidth{0pt}
\tablehead{
\colhead{Parameter} & \colhead{Units} & \colhead{Value} \tablenotemark{a}& \colhead{Source} 
}
\startdata
$T_{\textrm{eff,*}}$ & K & $3602 \pm 63$ & \cite{biddle:2013}\\
$\log_{10} g_*$ & (cgs) & $4.804 \pm 0.086$ & \cite{biddle:2013}\\

$P$ \tablenotemark{b}      & days & \porb  & \cite{biddle:2013}\\
$T_T$    & BJD$_{\tiny TDB}$ & 2\,456\,390.775\,80$\pm0.00040$ & This work \\
$i$      & deg  & $88.98^{+0.94}_{-1.25}$ & This work \\
$R_*/a$  & --   & 0.0702$^{+0.0076}_{-0.0027}$   & This work \\
$R_p/R_*$  & --  & 0.0789$^{+0.0021}_{-0.0019}$  & This work\\
$c$  & --  & $-0.351^{+0.025}_{-0.023}$  & This work\\
$d$  &  --  & $-0.889^{+0.051}_{-0.052}$  & This work\\
$\langle \beta_{T12} \rangle$ & -- & 1.34 & This work 
\enddata
\tablenotetext{a}{Uncertainties shown have been inflated by $\langle
  \beta_{T12} \rangle$}
 \tablenotetext{b}{Imposed as a Gaussian prior
  in the fitting process.}
\end{deluxetable}

\clearpage

\begin{deluxetable}{c c c c}
\tabletypesize{\scriptsize}
\tablecaption{  Limb-Darkening Priors (\S~\ref{sec:limbdarkening}) \label{tab:ldprior}}
\tablewidth{0pt}
\tablehead{
\colhead{Wavelengths} &\colhead{$c \pm \sigma_c$} & \colhead{$d \pm \sigma_d$} & \colhead{$\sigma^2_{cd}$}\\
\colhead{[$\mu m$]}  & \colhead{} & \colhead{} & \colhead{$[10^{-4}]$}
}

\startdata
 0.73--0.78 & $-0.266 \pm 0.043$ & $1.156 \pm 0.032$ & -7.84 \\
 0.78--0.82 & $-0.292 \pm 0.024$ & $1.124 \pm 0.024$ & 2.38 \\
 0.83--0.88 & $-0.347 \pm 0.018$ & $1.183 \pm 0.040$ & 3.63 \\
 0.88--0.92 & $-0.331 \pm 0.023$ & $1.115 \pm 0.028$ & -0.81 \\
 0.93--0.97 & $-0.330 \pm 0.024$ & $1.061 \pm 0.026$ & -2.38 \\
 0.97--1.02 & $-0.306 \pm 0.023$ & $0.994 \pm 0.019$ & -1.69 \\
 2.09--2.13 & $-0.389 \pm 0.032$ & $1.001 \pm 0.050$ & -13.84 \\
 2.13--2.17 & $-0.385 \pm 0.022$ & $0.971 \pm 0.039$ & -6.84 \\
 2.17--2.21 & $-0.361 \pm 0.017$ & $0.905 \pm 0.035$ & -4.54 \\
 2.21--2.26 & $-0.336 \pm 0.021$ & $0.843 \pm 0.037$ & -6.49 \\
 2.26--2.31 & $-0.306 \pm 0.024$ & $0.785 \pm 0.035$ & -6.94 \\
 2.31--2.36 & $-0.300 \pm 0.019$ & $0.794 \pm 0.028$ & -4.23 \\
\hline
0.73--1.02 & $-0.310 \pm 0.024$ & $1.124 \pm 0.022$ & 1.32 \\
2.09--2.36 & $-0.350 \pm 0.015$ & $0.892 \pm 0.031$ & -3.45 
\enddata

\end{deluxetable}
\clearpage

\begin{deluxetable}{c c c c c c c}
\tabletypesize{\scriptsize}
\tablecaption{  MOSFIRE Wavelength-Integrated (2.09--2.36\,\micron) Model Fits \label{tab:mosfirebic}}
\tablewidth{0pt}
\tablehead{
\colhead{State Vectors} & \colhead{SDNR} & \colhead{$\Delta \chi^2$} & \colhead{$\Delta$BIC}  & \colhead{$\langle \beta_{T12} \rangle$} & \colhead{$T_T - 2\,456\,391$ \tablenotemark{a} } & \colhead{$R_p/R_*$ \tablenotemark{a}} \\
\colhead{} & \colhead{[ppm]} & \colhead{}& \colhead{}& \colhead{}& \colhead{[BJD$_{\textrm{TDB}}$]}& \colhead{}
}

\startdata
$x$             &    1121 & 20.0 &  8.0 & 1.34 & -0.22420$^{+0.00030 }_{ -0.00031}$ &0.0789$^{+0.0014 }_{ -0.0013}$\\
$x,x^2$     &        1110  & 29.0 & 5.1 & 1.30 & -0.22432$ ^{+0.00030 }_{ -0.00033}$ & 0.0793$ ^{+0.0014 }_{ -0.0013}$\\
$w$         &        1128  & 15.3 & 3.3 & 1.27 & -0.22447$^{+0.00030}_{-0.00031}$   &  0.0788$^{+0.0014}_{-0.0015}$ \\
$x,w$      &        1113  & 25.4 & 1.5 & 1.28 &  -0.22426$^{+0.00030}_{-0.00028}$  & 0.0787$^{+0.0015}_{-0.0015}$\\
$w,wx$       &        1113 & 25.3  & 1.4 & 1.30 & -0.22428$^{+0.00032 }_{ -0.00030}$ &0.0783$^{+0.0015 }_{ -0.0014}$\\
$y,x,x^2$   &        1106  & 36.6 & 0.7 & 1.25 & -0.22416$ ^{+0.00040 }_{ -0.00033}$ & 0.0788$ ^{+0.0015 }_{ -0.0019}$\\
None      &        1150 &  0   &   0  & 1.41 & -0.22445$^{+0.00030}_{-0.00032}$ &  0.0793$^{+0.0016}_{-0.0014}$

\enddata
\tablenotetext{a}{The uncertainties listed here have not yet been
  multiplied by the factor $\langle \beta_{T12} \rangle$.}
\end{deluxetable}

\clearpage

\begin{deluxetable}{c c c c c c}
\tabletypesize{\scriptsize}
\tablecaption{  \gjb\ Transmission Spectrum \label{tab:specdat}}
\tablewidth{0pt}
\tablehead{
\colhead{Wavelengths} & SDNR& $\langle \beta_{T12} \rangle$ &\colhead{$R_p/R_*$ \tablenotemark{a}} & \colhead{$c$} & \colhead{$d$} \\
\colhead{[$\mu m$]} & [ppm] &  &\colhead{$[10^{-2}]$} & \colhead{} & \colhead{} 
}
\startdata
2.09--2.13 & 1418 & 1.67   & $7.75^{+0.33}_{-0.36}$  & $-0.395^{+0.074}_{-0.061}$  & $1.015^{+0.099}_{-0.108}$  \\
 2.13--2.17 & 1292 & 1.35   & $8.00^{+0.27}_{-0.22}$  & $-0.385^{+0.034}_{-0.042}$  & $0.969^{+0.068}_{-0.070}$  \\
 2.17--2.21 & 1298 & 1.47   & $8.07^{+0.27}_{-0.24}$  & $-0.362^{+0.034}_{-0.032}$  & $0.906^{+0.067}_{-0.071}$  \\
 2.21--2.26 & 1203 & 1.26   & $7.85^{+0.21}_{-0.20}$  & $-0.337^{+0.034}_{-0.036}$  & $0.834^{+0.061}_{-0.057}$  \\
 2.26--2.31 & 1349 & 1.17   & $7.93^{+0.22}_{-0.26}$  & $-0.305^{+0.036}_{-0.039}$  & $0.782^{+0.056}_{-0.051}$  \\
 2.31--2.36 & 1275 & 1.19   & $7.67^{+0.24}_{-0.21}$  & $-0.300^{+0.034}_{-0.029}$  & $0.794^{+0.042}_{-0.051}$  
\enddata
\tablenotetext{a}{The confidence intervals listed here refer to the
  statistical uncertainties after applying $\langle \beta_{T12}
  \rangle$, as discussed in \S~\ref{sec:erranal}.}
\end{deluxetable}

\clearpage

\begin{deluxetable}{l r r}
\tabletypesize{\scriptsize}
\tablecaption{Goodness of Fit for Atmospheric Models \label{tab:models}}
\tablewidth{0pt}
\tablehead{
\colhead{Model Name} & \colhead{$\chi^2$} & \colhead{$P(\chi^2)^a$}
}
\startdata
                    Solar  &  52.83  &  $8.00 \times 10^{-8}$  \\
         50$\times$ solar  &  34.45  &  $1.55 \times 10^{-4}$  \\
        200$\times$ solar  &  17.51  &  $6.38 \times 10^{-2}$  \\
              Solar, no C  &  17.31  &  $6.78 \times 10^{-2}$  \\
         50$\times$, no C  &  22.84  &  $1.13 \times 10^{-2}$  \\
        200$\times$, no C  &  18.92  &  $4.13 \times 10^{-2}$  \\
                     Flat  &  15.08  &  $1.29 \times 10^{-1}$  
\enddata
\tablenotetext{a}{Probabilities calculated for 11 measurements and 10
  degrees of freedom.}
\end{deluxetable}

\clearpage 

\appendix
\section{A Closer Look at MOSFIRE}
\label{sec:mosfire}
MOSFIRE was designed to obtain high-quality NIR spectroscopy and
photometry of relatively faint objects; extragalactic and, to a lesser
extent, brown dwarf science helped to define the instrument's
capabilities. The study of transiting exoplanets (or other
high-precision measurements of bright sources) did not drive
instrument requirements; nonetheless the instrument still works quite
well for this purpose so long as certain important steps are
followed.

\subsection{Nonuniform and non-repeatable slit widths}
\label{sec:slitwidth}
For several reasons, MOSFIRE's spectroscopic flat frames have more
structure than past users of milled masks may expect.  First, MOSFIRE
has no internal flat-field calibration source and so flat frames are
acquired by pointing at the telescope dome. Thus telluric absorption
lines are imprinted on the initial flat frames. To the extent that a
particular wavelength remains matched to the identical pixel
throughout the night, this effect is unimportant because it will
divide out. However, any motion of the target in the dispersion
direction (whether because of uncorrected flexure, pointing wander in
our large slits, or from other sources) may induce systematic,
wavelength- and spectral-motion-dependent flux offsets into the final
spectra.

A more complicated effect results from MOSFIRE's use of a custom
cryogenic configurable slit unit (CSU) rather than milled masks. The
CSU allows rapid reconfiguration of mask designs
\citep[see][]{mclean:2010,mclean:2012}. The CSU consists of 46 pairs
of actuated bars, which move across the focal plane to create either
46 slits (each 7'' wide) or a smaller number of longer slits. The
slit-facing end of each bar is equipped with an infrared-black knife
edge.

Each bar's motion is repeatable to roughly 0.02'' on the sky
\citep{mclean:2012}, a value which corresponds to roughly 3\% of the
width of a typical 0.7''-wide slit (such as we used to obtain our flat
field frames). Thus any movement of the CSU bars between science and
flat field observations induce variations in the line profile of
several percent. These variations manifest themselves as horizontal
banding in the images, with each band corresponding to one of the 46
CSU bar pairs.  To avoid
deleterious effects, we recommend obtaining dome flats for a given
mask immediately before or after that mask's science observations.

Furthermore, the slits produced by the CSU exhibit nonuniformities in
width, which we hypothesize result from slight misalignments and
imperfections of the knife edges at the end of each CSU slit bar.  We
have not investigated the repeatability of this second effect, but
estimate its amplitude to again be several percent.  If calibration
files taken with one slit mask are used to calibrate a second mask (as
in our case, where high flux levels prevented us from taking flat and
arc frames with our 10''-wide slits), the flat frame must be
normalized in the spatial direction (in addition to the standard
normalization in the spectral direction). We merely divide each
detector column by the median of all detector columns, but other
approaches may prove more effective.

--Observe calibrations frames immediately before or after science
frames for a given mask design. Otherwise, avoid placing a star near
the border between two CSU bars.

\subsection{Fringing}
\label{sec:fringing}
Our experience shows significant fringing in wide-slit MOSFIRE spectra
taken in the vicinity of telluric OH emission lines.  In K band the
fringes reach a maximum (slit-width-dependent) amplitude of roughly
4~ADU~s$^{-1}$~arcsec$^{-1}$; in general the amplitude, spatial
frequency, orientation, and phase of these fringes all vary both
spatially and temporally. We have observed similar effects in similar
observations taken with Subaru/MOIRCS, and we hypothesize that a
similar phenomenon of lower amplitude could explain the
worse-than-expected performance of \cite{bean:2011}'s
spectrophotometric light curve obtained with Magellan/MMIRS in the H
band (where OH emission is particularly strong). MOIRCS instrument
scientist Dr.~I. Tanaka informs us that his instrument's fringing was
eliminated by replacing MOIRCS' standard plane-parallel filters with
wedged substrates. This success suggests a possible mitigation
strategy for MOSFIRE.

These fringes cause severe difficulties in extracting high-quality
spectra and high-precision time series data when observing in the
fixed-position ``staring'' mode that has become common in infrared
photometry and spectroscopy of transiting exoplanets
\citep[e.g.,][]{croll:2011b,bean:2011}. We explored a number of
strategies to remove the fringes. Sinusoidal and wavelet fitting fail
because the fringe parameters vary and cannot be accurately measured
at the most important location: within the region of high stellar
flux. Scaling and shifting a template sky frame, or related techniques
using Principal Component Analysis, also fail, both because the fringe
pattern evolves over time and because of rapid and non-correlated
variations in the sky emission spectrum at different wavelengths.

We therefore conclude that nodding the telescope along the slit axis,
and subtracting frames obtained at nearby times, is the optimal
MOSFIRE observing strategy whenever smooth, well-behaved background
levels are desired.  So long as the nod cadence is less than a few
minutes, telluric emission remains relatively unchanged and the
fringing emission is well-subtracted. Nod subtraction also helps
mitigate the slit width nonuniformities discussed above, because any
residual signature of this effect will largely subtract out. The
primary disadvantage is the S/N penalty incurred because the
background flux effectively doubles for the standard Poisson
calculations. Despite this, for bright targets the classically computed
S/N expected from MOSFIRE spectroscopy still exceeds that expected
from 8\,m-class telescopes even in the absence of nodding.

\clearpage

\begin{figure}[tb!]
\centering
\includegraphics[width=17cm]{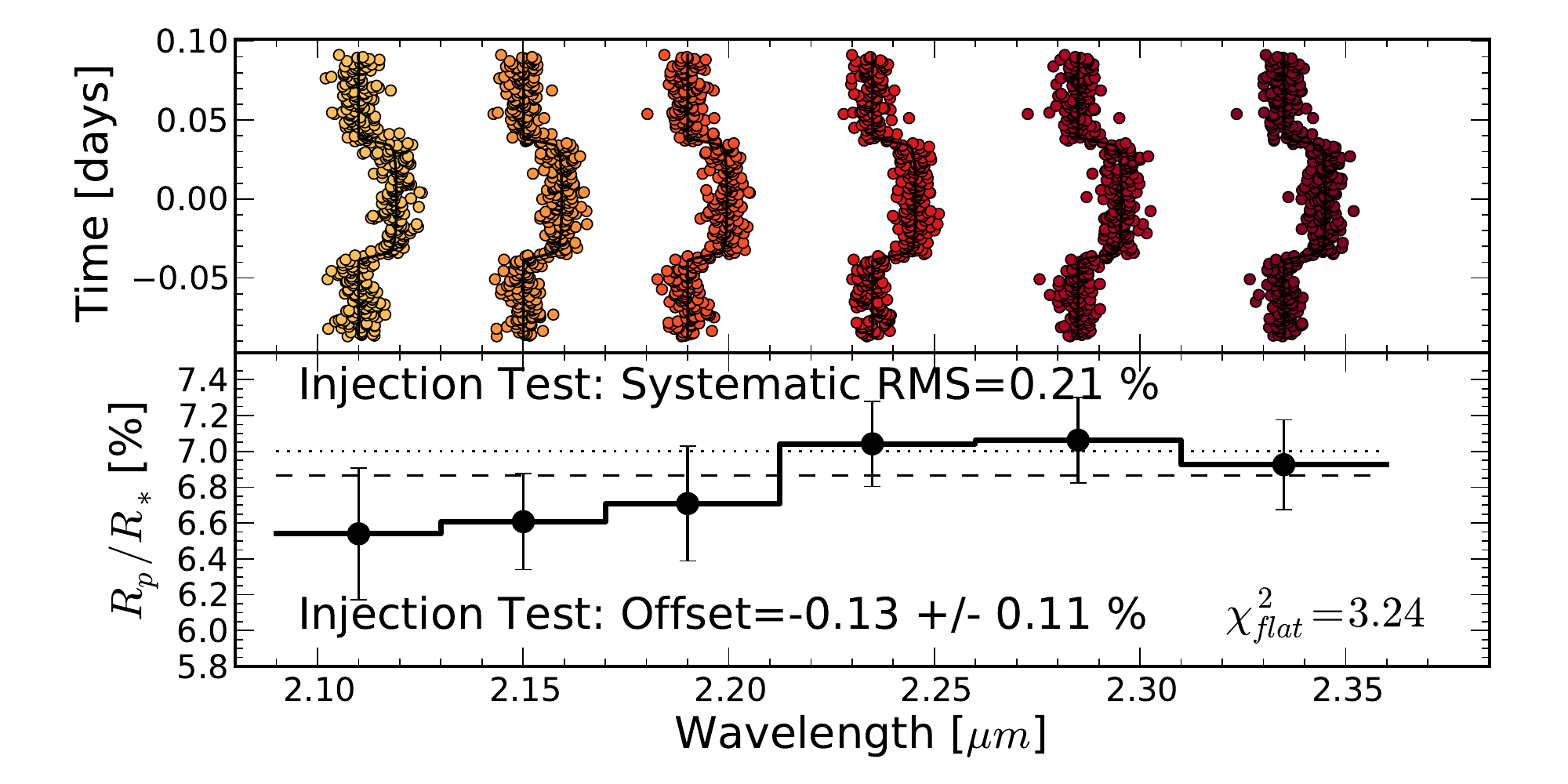}
\includegraphics[width=17cm]{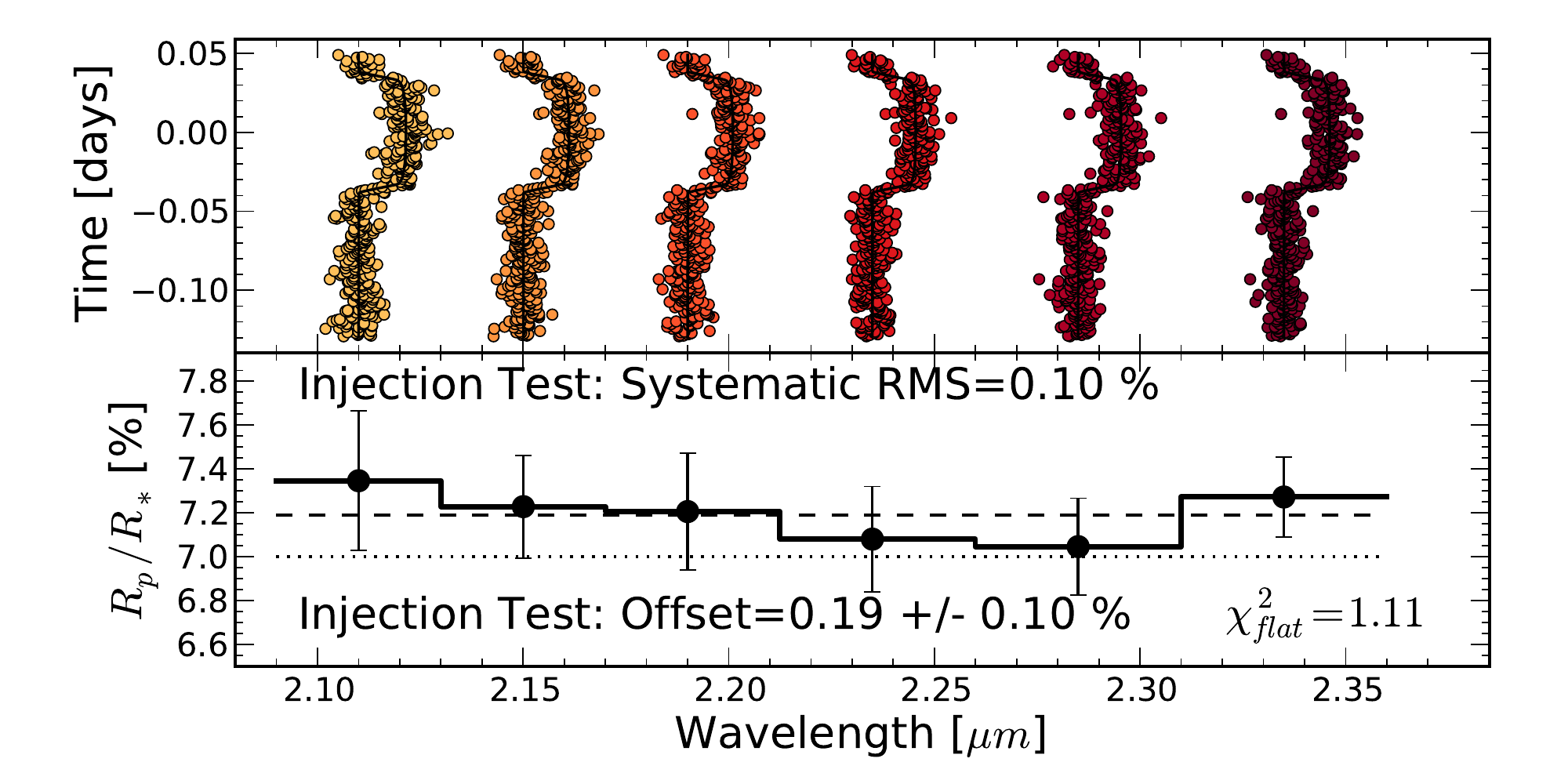}
\vspace{-0.2cm}
\caption{ \label{fig:mosinj} Injection-and-recovery results for K-band
  MOSFIRE data. The light curves shows {\bf simulated} transits that
  have been injected into real data and re-fit as described in
  \S~\ref{sec:injection}; they are plotted here after removal of
  overall trends and state vectors.  The dotted line shows the
  wavelength-independent injected value of 7.0\,\%. The thick solid
  lines and points shows the best-fit recovered transit parameters;
  the error bars derive from the MCMC analyses and have been inflated
  by $\langle \beta_{T14} \rangle$ as discussed in
  \S~\ref{sec:erranal}. The dashed lines indicates the weighted means
  for each test. In both cases the recovered spectra are consistent
  with a constant value, and the systematic offsets are $<2\sigma$.}
\end{figure}
\clearpage

\begin{figure}[tb!]
\centering
\includegraphics[width=17cm]{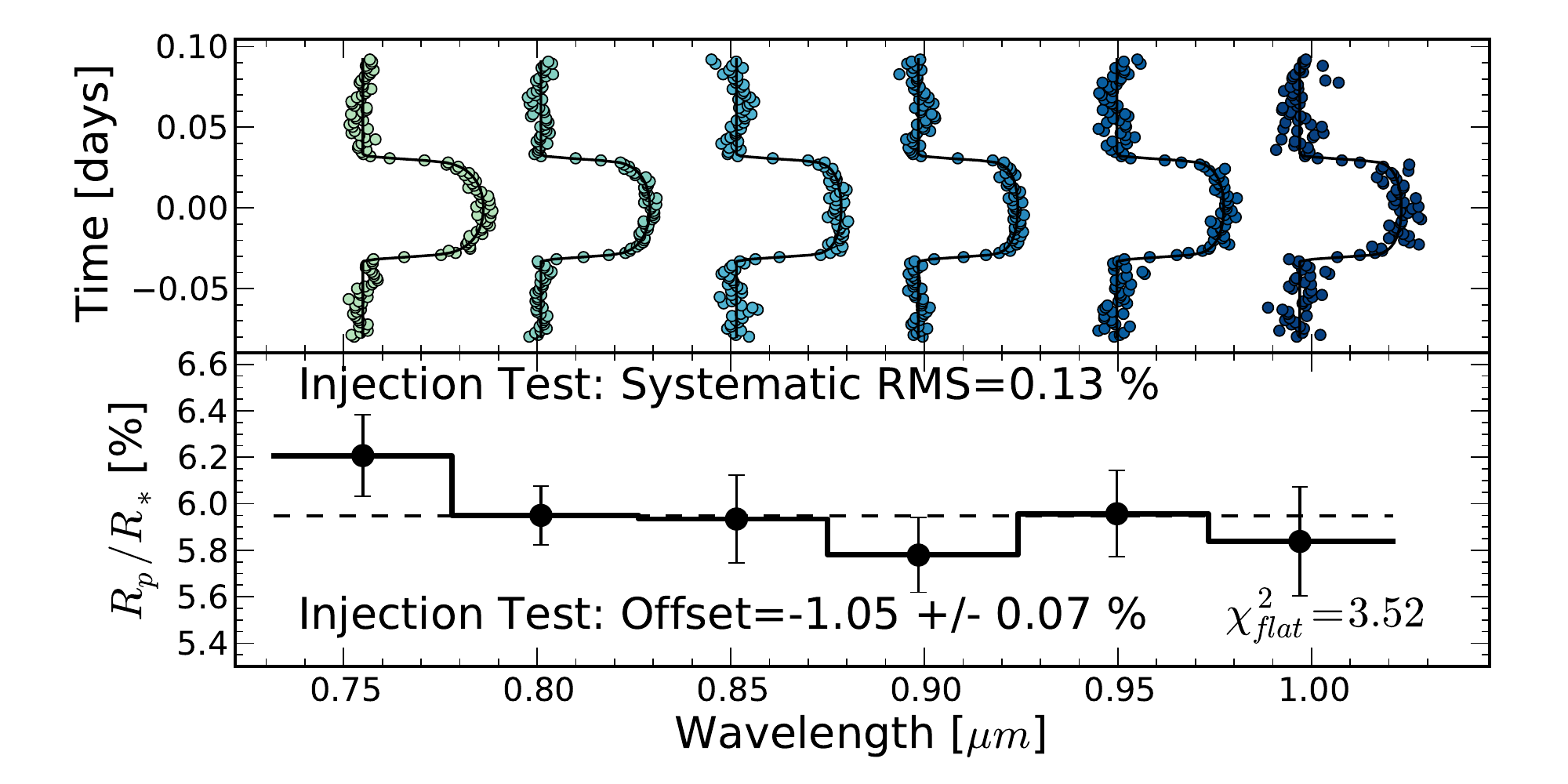}
\includegraphics[width=17cm]{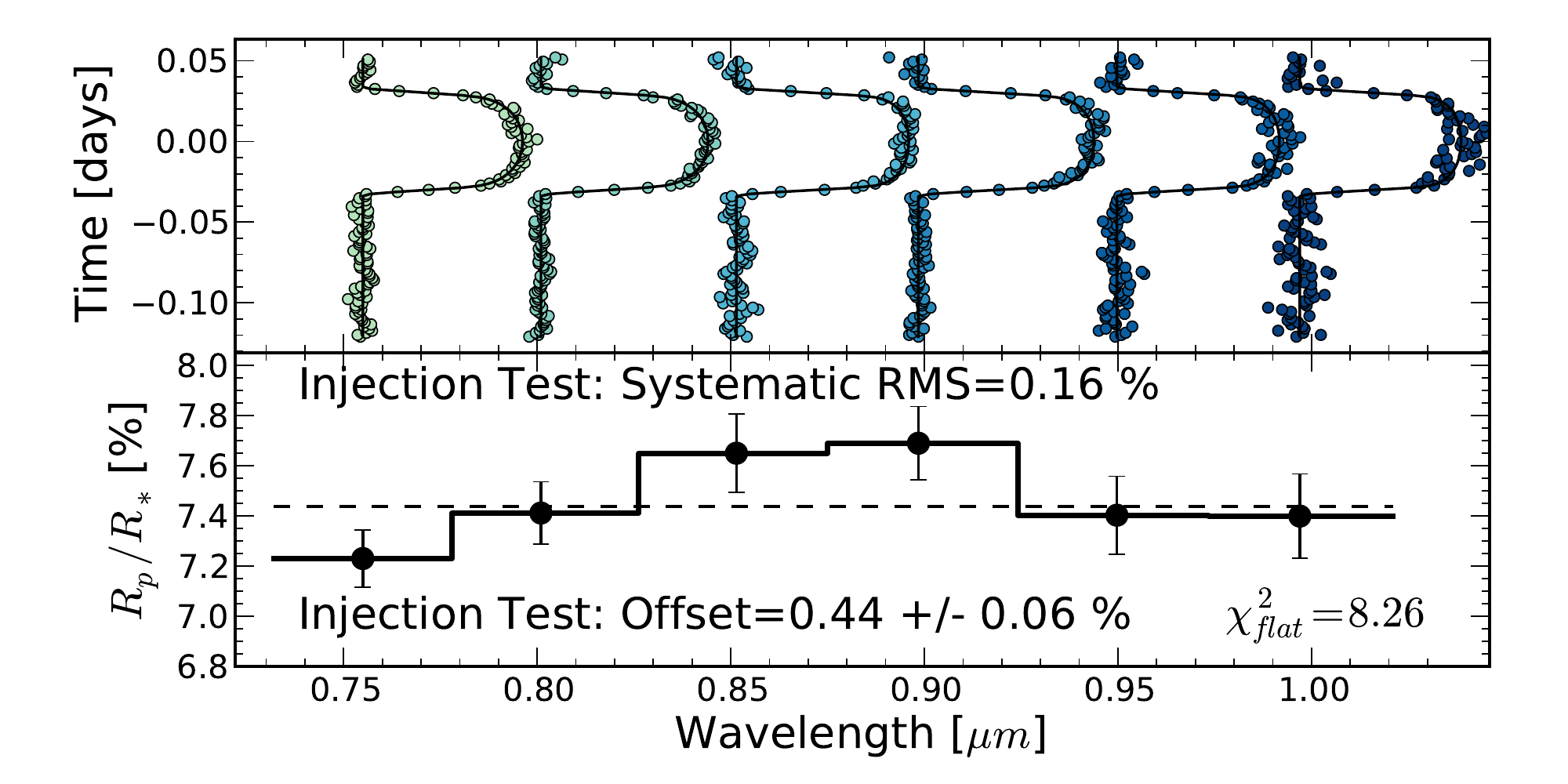}
\vspace{-0.3cm}
\caption{ \label{fig:gmosinj} Injection-and-recovery results showing
  {\bf simulated} light curves and results for optical GMOS data.
  Sub-figures and symbols are the same as Fig.~\ref{fig:mosinj}.
  Because we analyze the GMOS data in a differential fashion
  (\S~\ref{sec:whitelight}) these data are insensitive to the absolute
  transit depth. In addition, the retrieved spectrum in the bottom
  sub-figure exhibits considerable systematic errors and the systematics in
  the top sub-figure (though weaker) are similar in shape to the
  results of the true (non-injected) GMOS analysis shown in
  Fig.~\ref{fig:specfig}. }
\end{figure}
\clearpage
\clearpage

\footnotesize

\bibliographystyle{apj_hyperref}

\end{document}